\documentclass[prl,twocolumn,superscriptaddress,floatfix]{revtex4-2}
\usepackage{color}
\usepackage[english]{babel}
\usepackage{amsmath}
\usepackage{amssymb}
\usepackage{graphicx}
\usepackage[unicode=true,pdfusetitle,
 bookmarks=true,bookmarksnumbered=false,bookmarksopen=false,
 breaklinks=false,pdfborder={0 0 1},backref=false,colorlinks=true,urlcolor=blue,citecolor=blue,linkcolor=black]
 {hyperref}
\usepackage{subfigure}

\newcommand{\bsub}{\begin{subequations}}
\newcommand{\esub}{\end{subequations}}

\begin{document}
\title{Dissipative Hot-spot Enabled Shock and Bounce Dynamics \\ via Terahertz Quantum Quenches in Helical Edge States}
\def\rice{Department of Physics and Astronomy, Rice University, Houston, Texas
77005, USA}
\def\rcqm{Rice Center for Quantum Materials, Rice University, Houston, Texas
77005, USA}
\author{Xinghai Zhang}\affiliation{\rice}
\author{Matthew S.\ Foster}\affiliation{\rice}\affiliation{\rcqm}
\date{\today}

\begin{abstract}
We study quantum quenches of helical liquids with spin-flip inelastic scattering. 
Counterpropagating charge packets in helical edges can be created by an ultrashort electric pulse applied across 
a 2D topological insulator. Localized ``hot spots'' that form due to scattering enable two 
types of strongly nonlinear wave dynamics. First, propagating packets develop self-focusing shock fronts. 
Second, colliding packets with opposite charge can exhibit near-perfect retroreflection, despite strong dissipation. 
This leads to frequency doubling that could be detected experimentally from emitted terahertz radiation.
\end{abstract}
\maketitle

A small number of effective dynamical variables 
are typically
required to characterize equilibrium correlations 
in one-dimensional (1D) quantum systems \cite{GiamarchiBook,TsvelikBook,SutherlandBook}.
Strong driving out of equilibrium, as may be accomplished 
via a 
quantum quench \cite{QuenchReview}, can however excite an extensive number of variables. 
The ``solvable'' interactions
that govern 
1D
equilibrium physics are typically too kinematically constrained to relax such a system.
The most important dissipative effects are often induced by 
\emph{irrelevant operators}, which mediate momentum- and energy-mixing inelastic carrier-carrier collisions. 

Quantum dynamics relaxed through irrelevant interactions are difficult
to describe with otherwise powerful
tools such as quantum field theory \cite{ShankarBook}. A semiclassical alternative is 
hydrodynamics \cite{Landau} and its generalizations \cite{Bertini2016,Castro-Alvaredo2016}, 
which posit that rapid inelastic scattering scrambles quantum coherence, producing local equilibrium. 
At long times, only (nearly-) conserved hydrodynamic modes survive.

\begin{figure}[b]
\centering
\includegraphics[width=0.48\textwidth]{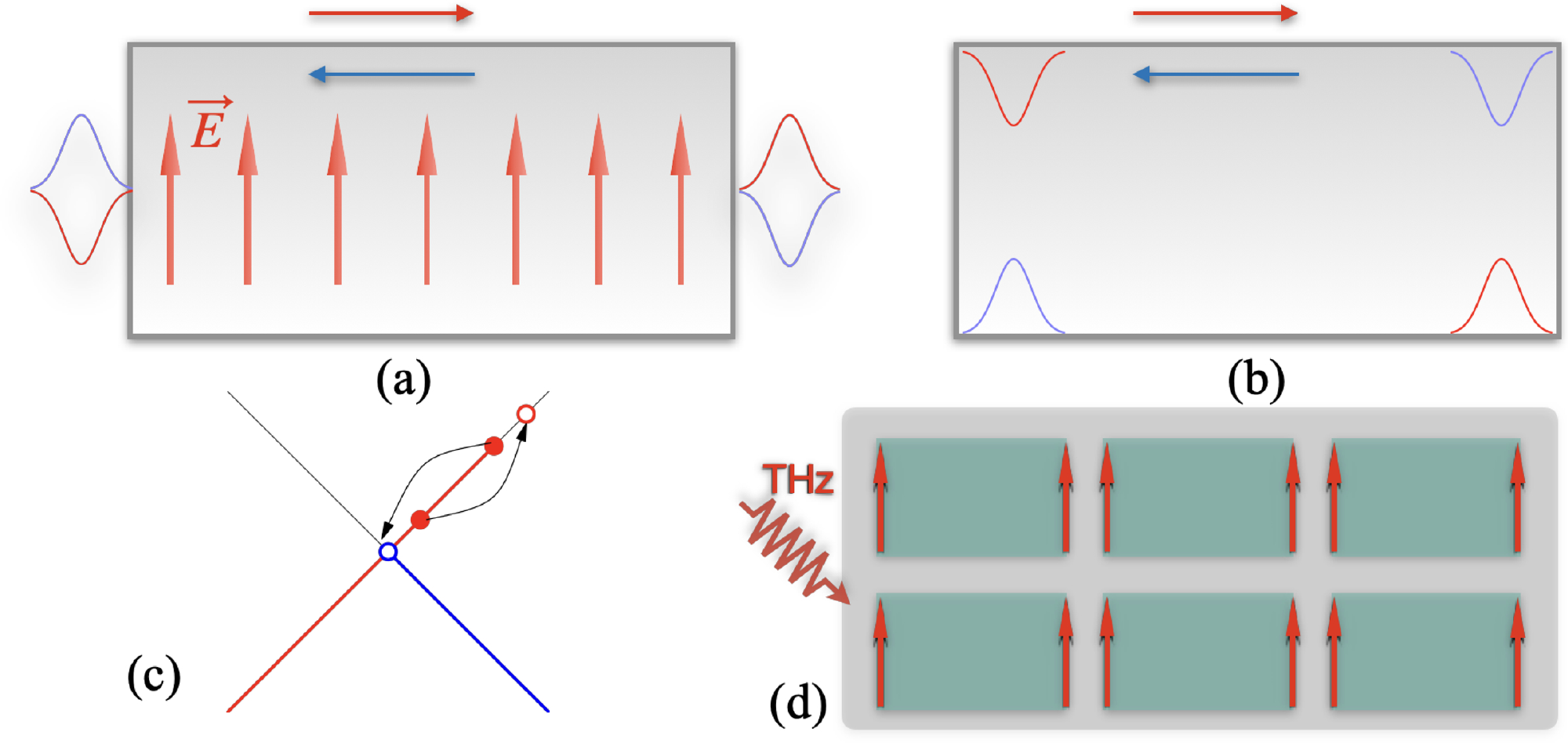}
\caption{
(a) An ultrashort laser pulse is applied to a rectangular 2D topological insulator (TI) sample, 
inducing equal and opposite charge density bumps of spin-up (red) and spin-down (blue) electrons on the left and right edges.
(b) The induced spin-up and -down electron bumps propagate in opposite directions, and round the corners to become 
isolated chiral packets on the long edges.
(c) The unusual 1PU scattering process: the momenta of two initial right movers is transferred
to a single right mover, while a left mover is emitted at zero momentum.
(d) An ``antenna'' formed by arrays of mesoscopic rectangular TI samples. 
The bumps induced by a laser pulse [as in (a)] produce time-periodic current switching along the field direction (indicated by red arrows). 
Strong 1PU scattering results in near-perfect bounce dynamics (see Fig.~\ref{fig:solitary}), leading to a frequency doubling of 
the current switching and emitted radiation, relative to the noninteracting case.
}
\label{fig:setup}
\end{figure}

Nonequilibrium physics is
of particular interest in topological edge 
states,
since these are well-understood to be ``topologically protected'' against interaction and disorder effects
in equilibrium \cite{Hasan2010,Qi2011}. 
In this Letter, we apply hydrodynamics to study
far-from equilibrium charge-packet dynamics in 1D helical liquids. 
Helical liquids can arise as edge states of two-dimensional (2D) 
$\mathbb{Z}_2$ topological insulators (TIs) \cite{Kane2005, Kane2005a, Bernevig2006, Konig2007},
hinge states of higher-order topological phases \cite{Schindler2018}, 
in graphene under strong magnetic fields \cite{Abanin2006, Fertig2006, Young2014}, 
and in quantum wires \cite{Quay2010, Oreg2010, Braunecker2009, Scheller2014, Kammhuber2017}. 

Although helical edge states have been reported in 2D TIs \cite{Konig2007,Du2015,Fei2017,Du2017,Reis2017,Wu2018,Lunczer2019},
the absence of ballistic conduction over large distances (compared to quantum Hall edge states)
remains a puzzle
\cite{Ma2015,Nichele2016,Dartiailh2020}.
Possible explanations include Kondo impurity scattering
\cite{Maciejko2009,Altshuler2013,Yevtushenko2015,Yevtushenko2020}, 
inelastic scattering due to electron-hole puddles \cite{Vayrynen2013,Vayrynen2014},
or interaction-mediated Anderson localization \cite{Wu2006,Xu2006,Hsu2017,Hsu2018,Chou2018}.
An interesting alternative invokes the breaking of axial spin symmetry due to Rashba spin-orbit coupling (RSOC)
\cite{Schmidt2012, Lezmy2012, Kainaris2014, Chou2015, Strunz2020}. RSOC enables an unusual, one-particle
umklapp (1PU) inelastic collision process 
[see Fig.~\ref{fig:setup}(c)], 
mediated by \emph{always-irrelevant} interactions.
In an ordinary quantum wire, 1PU scattering would be 
exponentially suppressed by incommensurability (away from the band edge).

In this Letter, we use chiral hydrodynamics to explore far-from-equilibrium
phenomena that could be induced in helical edges by a quantum quench,
mediated by 1PU scattering.
We show that an ultrashort electric-field pulse 
[such as a half-cycle terahertz (THz) laser pulse \cite{Shimano2013}]
applied across a 2D TI sample
can induce \emph{pairs} of oppositely charged right- and left-moving wavepackets on opposite 
edges, due to the chiral anomaly [see Fig.~\ref{fig:setup}(a)].
In the absence of RSOC, equal and opposite spin excesses 
are associated to the pairs formed at opposite sample edges.

\begin{figure}[t]
\centering
\includegraphics[width=0.49\textwidth]{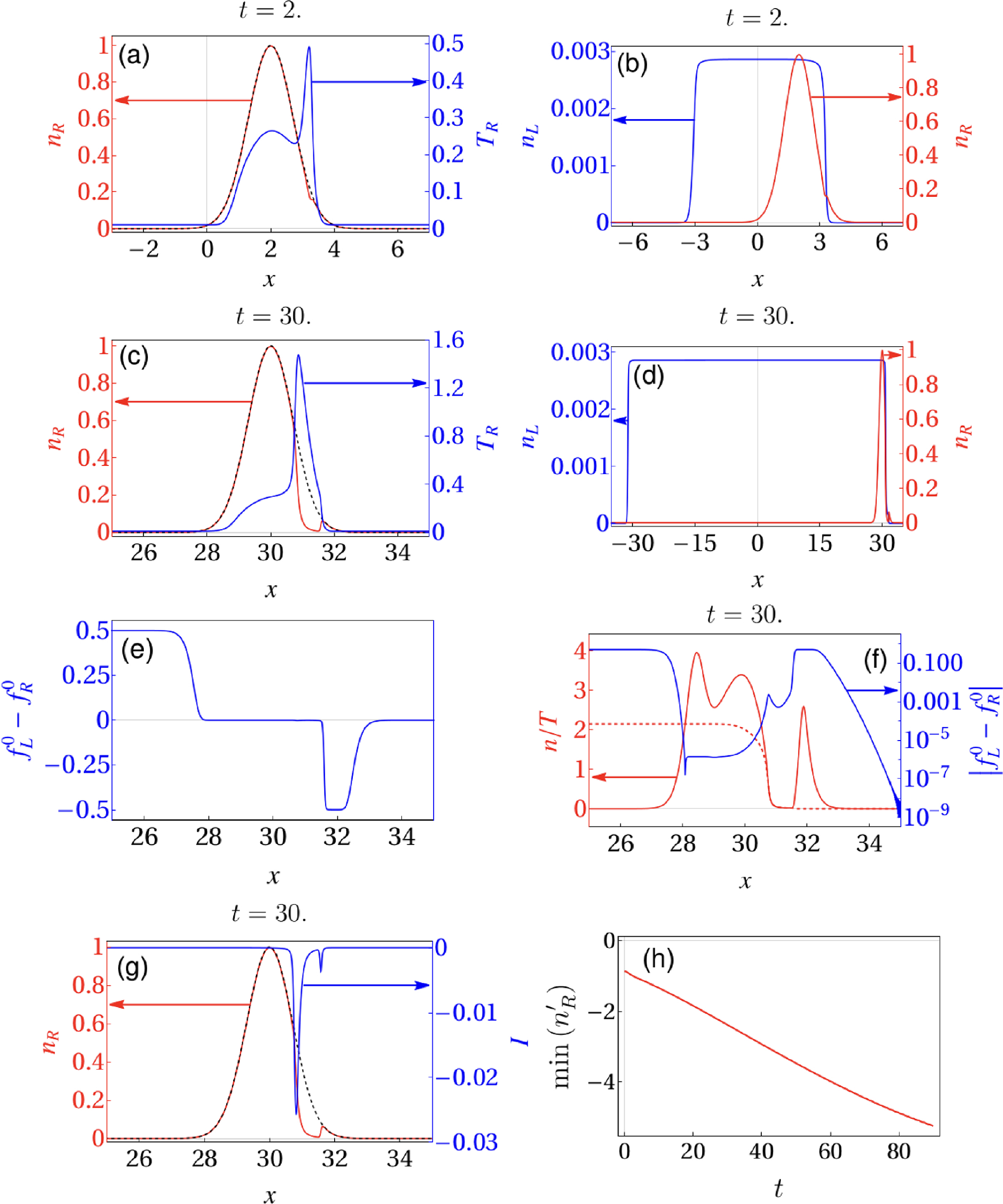}
\caption{Shock dynamics due to 1PU scattering for a right-propagating density packet. 
Here we set $v_F = k_B = 1$. 
The initial packet [Eq.~(\ref{eq:nR0})] has $n_0 = 1$ (which sets units for inverse length and energy) 
and 
$\xi = 1$, 
the initial temperature $T_0 = 0.01$,
and the interaction coupling $W = 2$.  
(a): Right-mover density $n_R$ (red curve) and temperature $T_R$ (blue curve) at time $t=2$. 
The dashed curve is the initial packet under a pure shift. 
(b): Right- (red) and left-mover (blue) density at $t=2$.
Note the small amplitude of the converted left-density,
which is close to its maximum value allowed by cooling $n_L = T_0/\sqrt{12} = 0.0029$ (see text).  
(c): Right-mover density and temperature at $t=30$.
The conversion to left movers at the hot leading edge produces a density shock. 
(d): Right- and left-mover densities at $t=30$. 
The $n_L$ amplitude is unchanged from (b). 
(e) The distribution function imbalance $f_L^0 - f_R^0$ at $t=30$.
$f_R^0 \sim 1$ from the start, since $n_0 \gg T_0$.  
After an initial transient, the left movers are cooled such that $f_L^0 - f_R^0$
is reduced to an exponentially small value in the bulk of the packet, leading to
negligible imbalance force $I$ there. 
(f): $n_R/T_R$(red curve), $n_L/T_L$ (red dashed curve) and $\left|f_L^0-f_R^0\right|$(blue curve).
(g): $n_R$ (red) and $I$ (blue) at $t=30$. 
(h): The slope of the right mover density at the shock wave front 
$\min{\left( n_R' \right)}$ versus $t$.}
\label{fig:nRL}
\end{figure}

Without inelastic carrier-carrier scattering, the field-induced pulses would 
ballistically propagate along the edges with rigid shape profiles [Fig.~\ref{fig:setup}(b)].
With 1PU scattering, the packets still propagate ballistically, but ``melt''
through shock formation, see Fig.~\ref{fig:nRL}.
We show that the 1PU shocks are intrinsically different from those arising 
due to band curvature in ordinary quantum wires
\cite{Bettelheim2006}: here, a density shock forms
due to a localized ``hot spot'' at the leading edge of a right-moving charge packet, where right movers
are steadily converted to left movers. Dissipation is confined to the hot spot, which 
enables quasiequilibration throughout the rest of the packet. 
The 1PU-mediated shock mechanism works at any nonzero temperature for doping close to the 
edge Dirac point. 

\begin{figure}[t]
\centering
\includegraphics[width=0.50\textwidth]{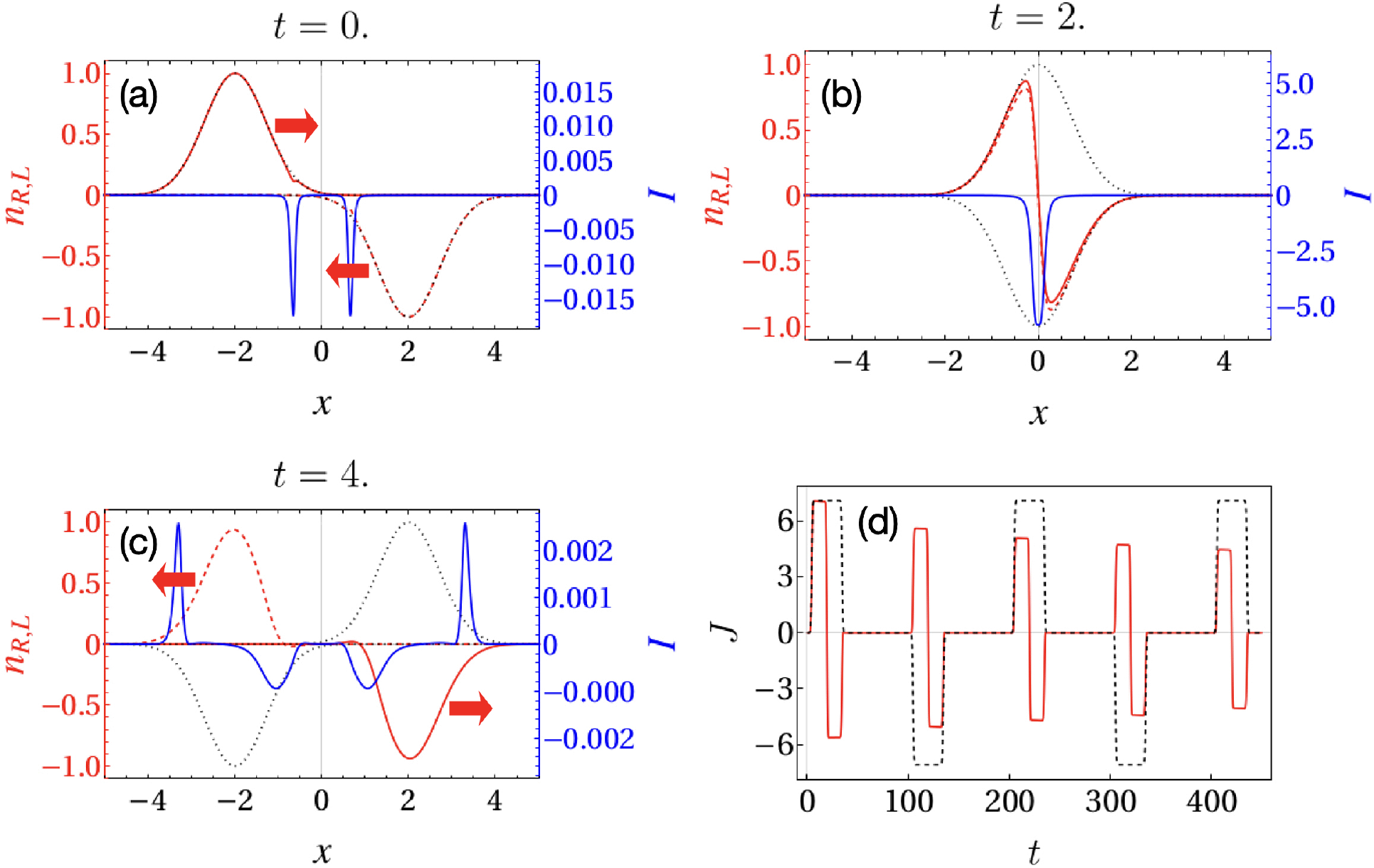}
\caption{Bounce-back dynamics for two colliding packets with opposite charge density. 
Here we have $n_0 = 1$, $\xi=1$, $T_0=0.001$, and $W=1$. 
(a): Density of right $n_R$ (red) and left $n_L$ (red dashed) movers and the imbalance $I$ (blue) at $t=0$, 
prior to the collision. 
The imbalance peaks around the hot shock fronts. The red arrows indicate the direction of propagation. 
For comparison, ballistic propagation of the initial packets (rigid shifts) are shown by the black dotted curve. 
(b): The same as (a) at the central collision time $t=2$; 
$I$ peaks around the hot collision center and is much larger than at the shock fronts (a). 
(c): The same as (a) at $t=4$, after the collision; 
$I$ again peaks around the new shock fronts. 
(d): The current $J$ along the initial field direction in an isolated closed edge loop (see Fig.~\ref{fig:setup}), 
which oscillates in time with a frequency that doubles with 1PU scattering (red), compared to 
free ballistic propagation (black dashed curve).}
\label{fig:solitary}
\end{figure}

We also show that counterpropagating, oppositely-charged packets can exhibit near-perfect
retroreflection (``bouncing'') upon a collision, as shown in Fig.~\ref{fig:solitary}. 
This also relies on the formation of a localized hot spot;
here, stronger scattering paradoxically enables more ideal bounces. 
We predict a frequency-doubling effect for strong 1PU scattering in a closed edge loop of 2D TI, due 
to current switching induced by the bounce dynamics, see Fig.~\ref{fig:solitary}(d). 
In an array of mesoscopic edge loops patterned from a 2D TI, this effect could be detected 
through emitted radiation [Fig.~\ref{fig:setup}(d)].


\textit{Chiral Hydrodynamics}.---A 1D system of spinless or helical
electrons can be described by the 
kinetic equations,
\begin{equation}\label{KinEq}
	\partial_{t}f_{R,L}\pm v_{F}\partial_{x}f_{R,L}+eE\partial_{k}f_{R,L}
	=
	\mathcal{I}_{R,L}\left[f_{R},f_{L}\right].
\end{equation}
Here $f_{R,L}\left(x,k\right)$ denotes the distribution functions of right- and left-moving electrons, 
$eE$ is the electric force ($e < 0$), 
and $\mathcal{I}_{R,L}$ are the collision integrals obtained from the Keldysh formalism \cite{Kamenev2011}.
Inelastic scattering mediated by irrelevant interactions can be encoded in $\mathcal{I}_{R,L}$.
Marginal Luttinger liquid interactions can be eliminated before deriving the kinetic equations, 
by bosonizing and refermionizing in terms of new fermion fields carrying fractional electric charge \cite{Imambekov12}; 
the price paid is induced nonlocality in $\mathcal{I}_{R,L}$.

Subtracting the zero-temperature filled Fermi seas below the helical edge Dirac point (defined to occur at $k = 0$) with 
$\tilde{f}_{R,L} \equiv f_{R,L} - \theta\left(\mp k\right)$, where $\theta(k)$ denotes the unit step function,
Eq.~(\ref{KinEq}) becomes 
\begin{equation}
\!\!
	\left(
		\partial_{t}
		\pm 
		v_{F}\partial_{x}
		+
		eE
		\partial_{k}
	\right)
	\tilde{f}_{R,L}
	\mp 
	eE \, \delta\left(k\right)
	=
	\mathcal{I}_{R,L}\left[f_{R},f_{L}\right].
\!\!
\end{equation}
Integrating these in momentum gives the chiral hydrodynamic equations
\begin{align}
	\partial_{t}n_{R,L}\pm v_{F}\partial_{x}n_{R,L} 
	=&\,
	\pm
	{eE}/({2\pi})
	\pm 
	I\,,
	\label{eq:nEq}
\\
	\partial_{t}P_{R,L}\pm v_{F}\partial_{x}P_{R,L} 
	=&\,
	eE
	\,
	n_{R,L}
	+
	F_{in}^{R,L}\,.
	\label{eq:PEq}
\end{align}
Here $n_{R,L}$ and $P_{R,L}$ respectively denote the particle and momentum densities
of right- and left-moving fermions (relative to the subtracted background). 
The source terms 
$
	I_{R,L}
	=
	\frac{1}{2\pi}
	\int_{-\infty}^{\infty}
	dk
	\,
	\mathcal{I}_{R,L}
	\equiv
	\pm 
	I
$
and 
$
	F_{in}^{R,L}
	\equiv
	\frac{1}{2\pi}
	\int_{-\infty}^{\infty}
	dk
	\,
	k
	\,
	\mathcal{I}_{R,L}
$
respectively describe the imbalance relaxation between right- and left-moving fermions 
and the friction forces.
The $\pm eE/2\pi$ term in Eq.~(\ref{eq:nEq}) is the chiral anomaly \cite{Nielsen1983,Son2013,Goswami2015}
(in units such that $\hbar = 1$).

With RSOC, the locked spin orientation of the helical carriers twists with increasing momentum 
\cite{Schmidt2012}, so that the density operator develops a backscattering component \cite{Xie2016}. 
As a result, screened Coulomb interactions induce a new one-particle umklapp (1PU) interaction 
between helical electrons. The full Hamiltonian for the helical edge reads
$
	H 
	=
	H_{0}	+	H_{\text{1PU}}, 
$
where
\begin{align}
	H_{0} 
	=&\,
	-i v_{F}
	\int dx
	\left(
		R^{\dagger} \partial_{x} R 
		-
		L^{\dagger} \partial_{x} L
	\right),
\\
	\!\!\!\!
	H_{\text{1PU}} 
	=&\,
	W
	\int dx
	\left[
		\begin{aligned}
			&\,
			R^{\dagger} R L^{\dagger}(-i\partial_{x}) R
		\\
			+
			&\,
			L^{\dagger} L R^{\dagger}(-i\partial_{x}) L
			+
			\text{H.c.}
		\end{aligned}
	\right]\!.
\end{align}
Here $H_{0}$ describes right-($R$) and left-($L$) moving free helical electrons, 
and $H_{\text{1PU}}$ is the 1PU interaction induced by Coulomb repulsion and RSOC. 
The coupling $W$ carries units of squared-length/time; 
this is the \emph{least} irrelevant operator that can relax a density imbalance. 

With local equilibrium in helical edge states, the distribution of particles obey the Fermi-Dirac distribution
\begin{equation}
	f(\varepsilon, T) 
	= 
	1/(e^{\tilde{\varepsilon}/{k_B T}}+1).
	\label{eq:Fermi-Dirac}
\end{equation}
Here 
$\tilde{\varepsilon} \equiv v_F k - \mu_R$ 
($\tilde{\varepsilon} \equiv -v_F k-\mu_L$)
for right (left) movers, with $\mu_{R,L}$ the chemical potentials. 
The right- and left-moving electrons can also take different temperatures. 
For massless relativistic 1D carriers, the hydrodynamic distribution function 
is fully parameterized by spatially-dependent temperatures and chemical potentials,
in contrast to the usual case with nonlinear dispersion, where the hydrodynamic (boost) 
velocity must also be introduced \cite{Levchenko2010,Levchenko2011}. 
An interesting feature of the 1PU interaction
is that its friction forces $F_{in}^{R,L}$ [Eq.~(\ref{eq:PEq})] 
both vanish in the clean limit: the right- and left-mover momenta are separately conserved,
a property shared by the more conventional 
two-particle umklapp (sine-Gordon \cite{GiamarchiBook}) interaction. 
The hydrodynamic equations with 1PU scattering can be benchmarked by
calculating thermoelectric transport coefficients \cite{SM}.


\textit{Shock Wave}.---Consider a helical edge realized at the boundary of a rectangular 
2D TI
sample. 
An ultrashort THz laser pulse can direct a transient electric field across the sample, 
see Fig.~\ref{fig:setup}(a). Due to the chiral anomaly [Eq.~(\ref{eq:nEq})], a half-cycle pulse 
will coherently convert a swath of left movers into right movers, creating pairs of 
equal and opposite left- and right-moving density bumps on opposite sample sides \cite{SM}. 
If the pulse duration is shorter than the 
inelastic relaxation time, the imbalance force $I$ can be neglected during this transfer
process, which is isothermal in this case. 
After the cessation of the field and assuming weak enough inelastic scattering for the left-
and right-moving bumps to 
separate and turn around the corners [Fig.~\ref{fig:setup}(b)], 
the subsequent evolution is determined by the 
clean hydrodynamic equations with $E = 0$. 
We assume that the system is disorder-free so that $F_{in}^{R,L} = 0$ in Eq.~(\ref{eq:PEq}), 
with an initial (pre-pulse) equilibrium state
doped to the Dirac point ($n_R = n_L = 0$), 
at low but nonzero temperature $T_R = T_L = T_0$.

We will focus on the effects of inelastic scattering for the right propagation of the 
excess right-mover density $n_R(t,x)$ induced by the field. 
Shortly after the splitting with the left-moving density deficit, we assume that the right-moving excess
has the initial shape
\begin{equation}
	n_R^\text{\tiny (0)}(x) 
	= 
	n_0 \exp\left[-\left(x/\xi\right)^2\right].
  \label{eq:nR0}
\end{equation}
The total excess $N = \sqrt{\pi} n_0 \xi$ is set by the product of the peak laser pulse amplitude 
and duration. 
The width $\xi$ is determined by the length of the transverse edge of the sample 
and the pulse duration of the laser \cite{SM}.

In what follows, we set $v_F = k_B = 1$. 
The 1PU scattering process can transfer the momenta of two initial right-moving electrons
to a single right mover, whilst simultaneously emitting a left mover at the Dirac point with zero momentum, 
see Fig.~\ref{fig:setup}(c). For the right propagation of the right-mover excess density $n_R(t,x)$, 
it means that right movers are effectively \emph{heated} while
left movers are effectively \emph{cooled}. In particular, since $P_{R,L}$ are separately conserved, 
we have 
\begin{align}\label{eq:PRL}
	P_{R,L}(t,x) 
	= 
	\pm\pi
	\left\{ 
		{T_0^2}/{12} 
		+ 
		\left[ n_{R,L}^\text{\tiny (0)}(x_{\mp}) \right]^2 
	\right\},
\end{align}
using the equilibrium expression for $P_{R,L}$ in terms of $T_{R,L}$ and $n_{R,L}$ \cite{SM}. 
Assuming local equilibration, the same relation can be used to write $T_{R,L}$ in terms of the densities
and momenta, leading to 
\begin{align}\label{eq:TRL}
\!\!\!
	T_{R,L}(t,x)
	=&\, 
	\sqrt{ 
		T_0^2 
		+ 
		12 \left\{ 
			\left[ n_{R,L}^\text{\tiny (0)}(x_{\mp}) \right]^2 
			-  
			\left[n_{R,L}(t,x)\right]^2 
		\right\}
	}.
\!\!\!
\end{align}
In Eqs.~(\ref{eq:PRL}) and (\ref{eq:TRL}), $n_{L}^\text{\tiny (0)} = 0$. 
Here we have introduced lightcone coordinates  
$x_\pm = x \pm t$ and $\partial_\pm = \pm \frac{1}{2}(\partial_t \pm \partial_x)$.
The density dynamics are determined by [Eq.~(\ref{eq:nEq})] 
\begin{equation}
	2
	\,
	\partial_{\pm} 
	n_\text{R,L}\left(x_{+}, x_{-}\right)
	=
	I(n_R,n_L,T_R,T_L),
	\label{eq:partial+-}
\end{equation}
where for the 1PU mechanism \cite{SM}
\begin{equation}
	\!\!\!\!
	I 
	= 
	\frac{\pi^2 W^2}{96}
	\left(f_L^0 - f_R^0\right)
	\!
	\left[
		\begin{aligned}
			\left(4 n_R^2 + T_R^2\right)
			\left(4 n_R^2 + 9 T_R^2\right)
		\\ 
			+
			\left(4 n_L^2 + T_L^2\right)
			\left(4 n_L^2 + 9 T_L^2\right)
		\end{aligned}
	\right]\!.
	\!\!
\label{eq:I}
\end{equation}
In this equation, $f_{R,L}^0 \equiv \left[e^{-2 \pi n_{R,L} / T_{R,L}} + 1\right]^{-1}$. 

In a similar setup for (e.g.) a spinless wire with 2-particle umklapp (2PU) 
scattering \cite{GiamarchiBook},
Eqs.~(\ref{eq:PRL})--(\ref{eq:partial+-}) would also apply. 
A crucial difference, however, arises in the expression for $I$ in Eq.~(\ref{eq:I}).
Since the 1PU process upconverts the momenta of excess right movers while emitting
left movers only at zero momentum, this can work at arbitrarily low but
nonzero temperature for a system doped to the edge Dirac point. 
We assume the far-from-equilibrium limit where the initial excess right-density $n_0 \gg T_0$. 
In 2PU scattering, two right movers are converted to two left movers
with equal and opposite momenta. At low temperature for a system nominally doped
to the Dirac point, such processes would be exponentially suppressed by Pauli-blocking of the left movers.

To determine the evolution of the density, we numerically integrate Eq.~(\ref{eq:partial+-}),
using Eqs.~(\ref{eq:TRL}) and (\ref{eq:I}). We work with a periodic system of length $L$, 
and solve the resulting coupled equations for the momentum modes $n_{R,L}(t,k)$ \cite{SM}. 
Key features of the dynamics are explained below; additional insight obtains
from approximate analytical calculations detailed in Ref.~\cite{SM}.

For $n_0 \gg T_0$, 
the right-moving pulse retains its initial shape
[Eq.~(\ref{eq:nR0})], except for the formation of a shock wave front, shown in Figs.~\ref{fig:nRL}(a,c,g).
This can be understood as follows. 
Initially $I$ in Eq.~(\ref{eq:I}) is large and negative throughout the pulse.
Immediately this induces the conversion of right to left movers, heating the former
and cooling the latter [Eq.~(\ref{eq:TRL})]; only this initial transient is captured by perturbation theory \cite{SM}.
The amplitude of the emitted left movers is limited by this cooling process,
since $n_L = T_0/\sqrt{12}$ would correspond to cooling to absolute zero. 
For $n_0 \gg T_0$, independent of $W$ this leads to a ``quasi-equilibration''
throughout the bulk of the packet, wherein 
$f_L^0 \rightarrow f_R^0 \lesssim 1$ suppresses the imbalance force $I$,
see Figs.~\ref{fig:nRL}(e,f). In order to sustain this balance, however,
left movers must be continuously converted from the leading (right) edge
of the packet, and this produces the density shock. 
The shock front is characterized by a spike in $T_R$ resulting from rapid heating, 
as shown in Figs.~\ref{fig:nRL}(a,c). 
The imbalance force $I$ is non-negligible only at the shock front [Fig.~\ref{fig:nRL}(g)], 
where $f_L^0-f_R^0$ is small but nonzero and $T_R$ peaks, and at the small kink in 
front of that, where $f_L^0 - f_R^0$ reaches a minimum negative value and $T_R$ and
$n_R$ are not too small.
The imbalance is positive but very small in the tail of emitted left movers.

\begin{figure}[t]
\centering
\includegraphics[width=0.45\textwidth]{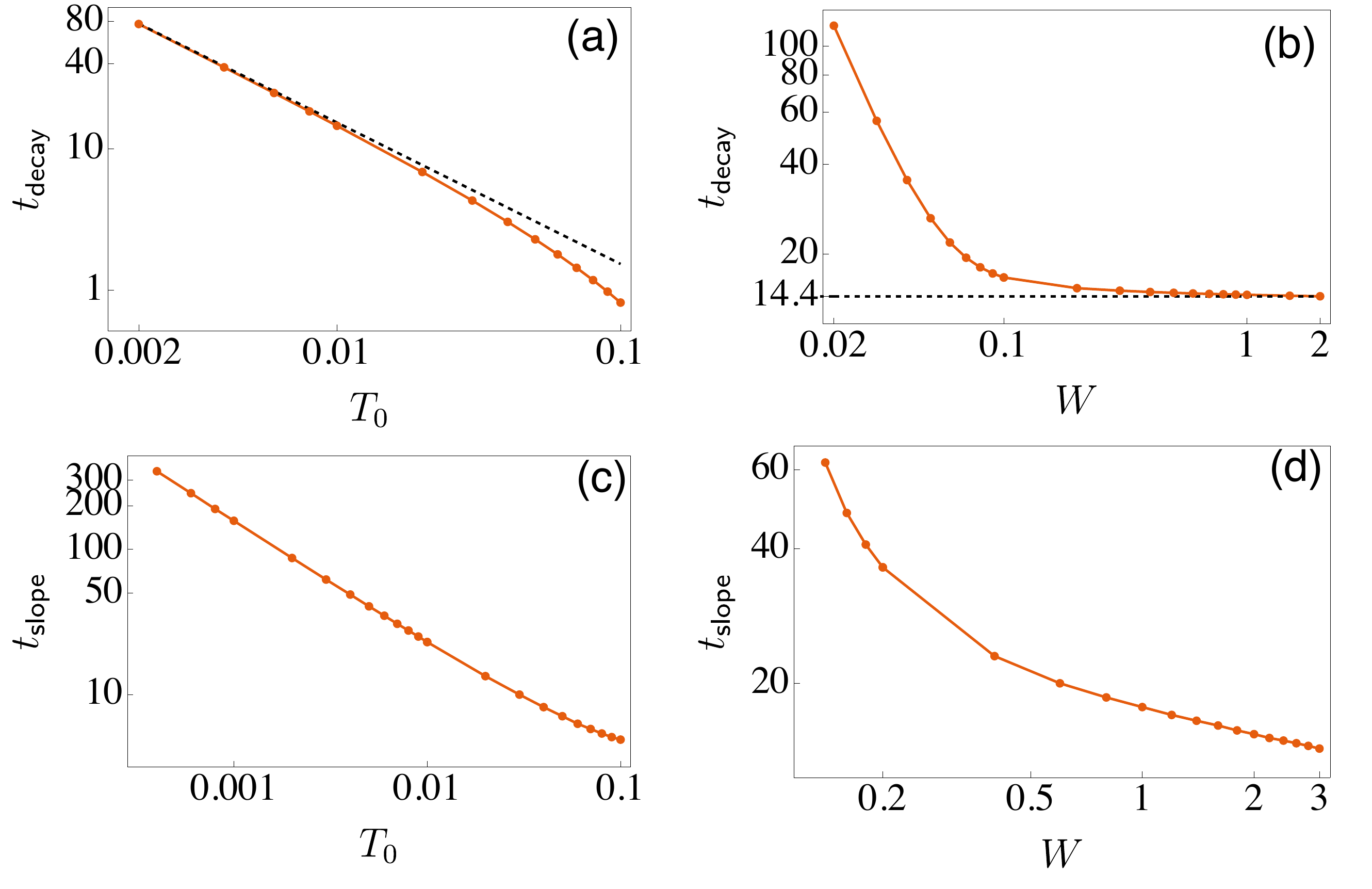}
\caption{Timescales for the shock dynamics, defined via Eq.~(\ref{tScales}).
(a): $t_\textsf{decay} \simeq 1/T_0$ at low temperature. The dashed line is the bound [Eq.~(\ref{eq:tdecay1})].
(b): $t_\textsf{decay}\sim W^{-2}$ for weak interactions, but saturates for $W \geq 0.1$. 
(c): $t_\textsf{slope} \sim T^{-0.86}$ at low temperature. 
(d): $t_\textsf{slope}$ decreases with smaller slope as $W$ increases, 
but does not saturate at strong interaction. 
Here $n_0=1$ and $\xi=1$ in all the plots, 
and $W=1.0$ in (a) and (c), 
whilst $T_0=0.01$ in (b) and (d).}
\label{fig:tdecay}
\end{figure}

The negative slope of the right mover density at the shock front decreases approximately linear in time $t$, 
as shown in Fig.~\ref{fig:nRL}(h). Different from the shock wave developed
in the inviscid Burgers' equation due to the nonlinear spectrum \cite{Landau,Bettelheim2006}, 
the shock front here arises entirely due to the 1PU scattering between particles 
and only reaches infinite slope at infinite time. 

To characterize the dynamics, we define the decay time $t_\textsf{decay}$ and the slope time $t_\textsf{slope}$ via
\begin{subequations}\label{tScales}
\begin{align}
	\Delta N_R(t) / N
	=&\,
	\alpha 
	\, 
	t / t_\textsf{decay},
\\
	\min{n'_R(t)} 
	\simeq&\,
	- 
	\left(n_0/\xi\right)
	\left(
		1
		+
		t / t_\textsf{slope}
	\right),
	\label{eq:tdecay}
\end{align}
\end{subequations}
where $N$ ($N_R$) is the initial (evolving) total right-excess and we set $\alpha=0.05$ in numerics.
The behavior of $t_\textsf{decay}$ and $t_\textsf{slope}$ are well-defined, 
since $N_R(t)$ and $\min\left[n'_R(t)\right]$ both exhibit linear dependence on time $t$
(after the initial transient quasi-equilibration behind the front).

As is shown in Fig.~\ref{fig:tdecay}(b), $t_\textsf{decay}$ is proportional to $W^{-2}$ for weak interactions,
but saturates to a fixed value for $W \gtrsim 0.1$. (Here we set $n_0 = 1$ to fix units.) 
If we assume that the emitted left movers are cooled to absolute zero with amplitude $n_L = T_0/\sqrt{12}$ 
[see Eq.~(\ref{eq:TRL})], then $t_\textsf{decay}$ asymptotes to 
\begin{equation}
	t_\textsf{decay} 
	= 	
	\sqrt{3} \alpha N /T_0,
\label{eq:tdecay1}
\end{equation}
valid in the long-time limit. 
This bound is nearly saturated for small $T_0 \ll n_0$ or $W \gg 1/n_0$, as shown in Figs.~\ref{fig:tdecay}(a,b). 
By contrast $t_\textsf{slope} \sim T_0^{-0.86}$, 
and does not saturate at strong interactions. 
Therefore, the shock front can form faster and faster for stronger interactions.

\textit{Bounces and Current Switching}.---Now we turn to interpacket collisions. 
When the density of the two packets share the same sign, they pass through each other nearly without scattering, since 
$I$ [Eq.~(\ref{eq:I})] is strongly suppressed for (e.g.) $n_{R,L} \gg T$. On the other hand, when two packets with opposite 
density collide, they \emph{bounce} back from each other for strong 1PU interaction, 
similar to solitary waves, see Fig.~\ref{fig:solitary}. The retroreflection becomes more perfect with lower initial 
$T_0/n_0$ 
and 
larger $W n_0$, i.e.\ larger interaction strength or initial density. 

When the packets move separately, the 
imbalance $I$ peaks around the hot shock fronts and local equilibration is maintained throughout the bulk of the packets [Fig.~\ref{fig:solitary}(a)]. 
When the two packets collide with each other, the density of the collision center goes close to zero and it heats up further. The collision center 
becomes the hot spot 
and $I$ becomes much larger than the case of shock front [Fig.~\ref{fig:solitary}(b)]. 
Around the collision center, the conversion of right movers to left movers is very efficient for $W n_0 \gtrsim 0.1$ \cite{SM}, 
leading to almost perfect retroreflection of the two packets. Then the two packets move apart and local equilibration again builds up 
throughout the packets except for the new shock fronts [Fig.~\ref{fig:solitary}(c)]. 
The bounce-back dynamics of the packets leads to frequency doubling of the current oscillation in a closed edge loop, compared 
to the case without 1PU scattering, as shown in Fig.~\ref{fig:solitary}(d). The slow decay of the current for strong interactions 
$W n_0 \gg 1$ is due to the shock formation mechanism between collisions, which operates at all $T_0 > 0$.

\textit{Conclusion}.---The dynamics predicted here could be realized in a quantum quench experiment,
applying an ultrashort electric field pulse across a 2D TI sample. 
Frequency-doubled radiation could be detected from an antenna 
formed by arrays of TI mesas, Fig.~\ref{fig:setup}(d). 
An 
interesting question is whether quantum effects modify the semiclassical predictions articulated here. 

\acknowledgments

We thank Alex Levchenko, Eldad Bettelheim, Emil Yuzbashyan, Yang-Zhi Chou, Douglas Natelson, and Keji Lai 
for helpful discussions. This work was supported by the Welch Foundation Grant No.~C-1809 and by NSF CAREER Grant No.~DMR-1552327.


\end{document}


\title{Supplemental Material: Dissipative Hot-spot Enabled Shock and Bounce Dynamics via Terahertz Quantum Quenches in Helical Edge States}
\author{Xinghai Zhang}
\affiliation{Department of Physics and Astronomy, Rice University, Houston, Texas
77005, USA}
\author{Matthew S.\ Foster}
\affiliation{Department of Physics and Astronomy, Rice University, Houston, Texas
77005, USA}
\affiliation{Rice Center for Quantum Materials, Rice University, Houston, Texas
77005, USA}
\maketitle
\tableofcontents

\section{Hydrodynamic Equations}

The helical edge liquid with 
one-particle umklapp (1PU) 
scattering defined by Eqs.~(5) and (6) in the main text is described by the following real-time action,
\begin{align}
\label{SRT}
	S
	=
	\intop_{t,x}
	\left\{ 
		\bar{R}\left[
				i \partial_{t}
				+
				i v_{F} \partial_{x}
				-
				V(x)
			\right]
		R
		+
		\bar{L}
			\left[
				i\partial_{t}
				-
				i v_{F} \partial_{x}
				-
				V(x)
			\right]
		L
	\right\} 
	-
	W
	\intop_{t,x}
	\left[
		\bar{R} R \bar{L} \left(-i\partial_{x}R\right)
		+
		\bar{L}L\bar{R}\left(-i\partial_{x}L\right)
		+
		\textrm{H.c.}
		\right].
\end{align}
In this action, we also include potential scattering;
here $V(x)$ denotes a short-range-correlated disorder potential with the Gaussian distribution,
\begin{equation}
	\overline{V(x) \, V(x')}
	=
	\lambda^{2}\delta\left(x-x'\right),
\end{equation}
where the overline denotes an average over disorder configurations. 
We perform the axial gauge transformation,
\begin{equation}
	R(x) 
	\to 
	e^{	-
		\frac{i}{v_{F}}
		\int^{x}dx' \, V(x')
	}
	R,
\qquad 
	L(x)
	\to 
	e^{
		\frac{i}{v_{F}}
		\intop^{x}dx' \, V(x')
	}L,
\end{equation}
leading to 
\begin{align}
	S
	\to 
	&\intop_{t,x}
	\left[
		\bar{R}\left(i\partial_{t}+iv_{F}\partial_{x}\right)R
		+
		\bar{L}\left(i\partial_{t}-iv_{F}\partial_{x}\right)L
	\right]
\nonumber\\
	&
	-
	W
	\intop_{t,x}
	\left[
		e^{-\frac{2i}{v_{F}}\int^{x}dx' \, V(x')}
		\,	
		\bar{R}R\bar{L}\left(-i\partial_{x}R\right)
		+
		e^{\frac{2i}{v_{F}}\int^{x}dx' \, V(x')}
		\bar{L}L\bar{R}\left(-i\partial_{x}\right)L
		+
		\textrm{H.c.}\right].
\end{align}
We ignore purely marginal Luttinger-liquid (current-current) interactions. In principle, these can be dealt with by 
bosonizing and subsequently refermionizing the system in terms of new fermion fields \cite{SM--GiamarchiBook}. 
When the latter are used to rewrite the action, the current-current interaction disappears, but the electric charge and Fermi velocity are
renormalized. The price paid is that the 1PU interactions, whilst still irrelevant, take on a nonlocal character
involving ``string operators'' \cite{SM--Imambekov12}. Since the new quasiparticles are otherwise noninteracting, one can still 
calculate a Fermi's golden rule rate for scattering, but we will not do this here. 

\begin{figure}[t]
  \centering
  \includegraphics[width=0.4\textwidth]{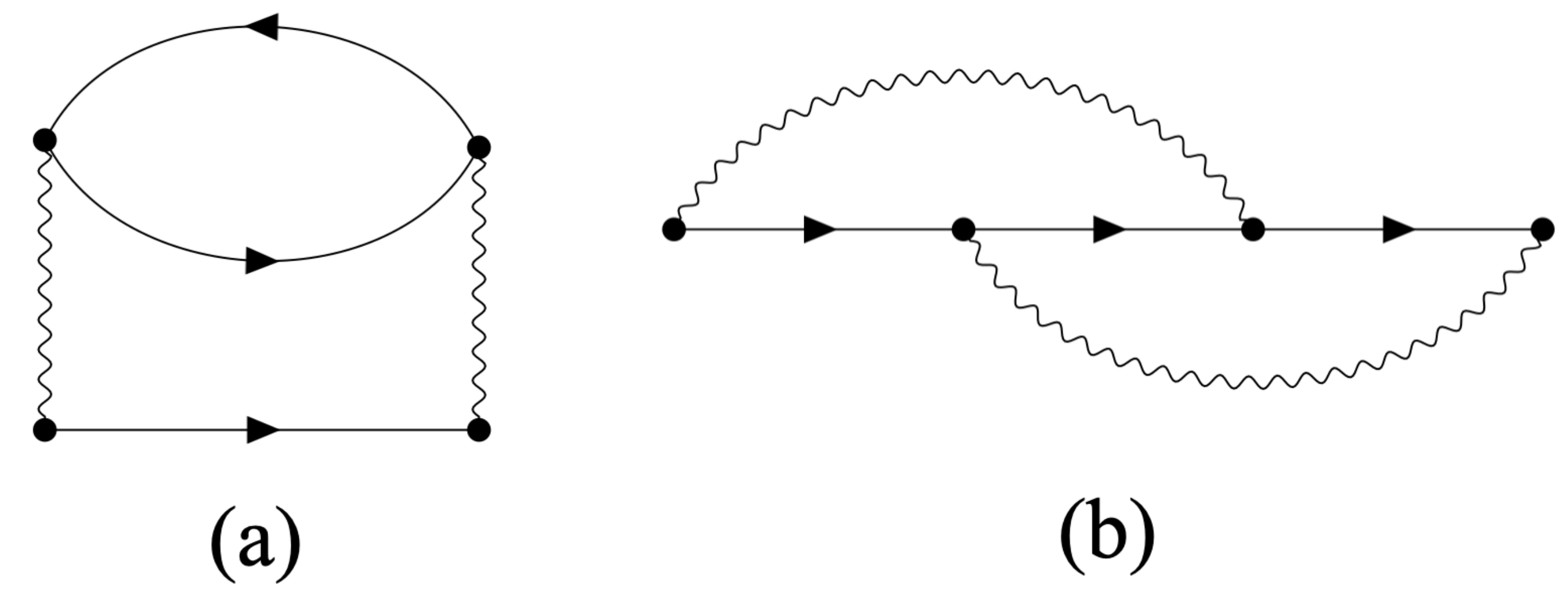}
  \label{fig:sigma}
  \caption{
	Non-equilibrium self-energies contributing to the collision integrals $\mathcal{I}_{R,L}$ in Eq.~(\ref{KEqs}). 
	The fermion internal lines are propagators for right- or left-moving fermions. 
	These lines also carry Keldysh labels.}
\end{figure}

In the quasi-particle approximation, we can use the Keldysh version of Eq.~(\ref{SRT}) to derive the following 
quantum kinetic equations for right- and left-moving electrons \cite{SM--Kamenev2011},
\begin{subequations}\label{KEqs}
\begin{align}
	\partial_{t}f_{R}+v_{F}\partial_{x}f_{R}+F\partial_{k}f_{R} & =\mathcal{I}_{R}\left[f\right]\,,\\
	\partial_{t}f_{L}-v_{F}\partial_{x}f_{L}+F\partial_{k}f_{L} & =\mathcal{I}_{L}\left[f\right]\,.
\end{align}
\end{subequations}
Here $F=eE$ is the electromagnetic force and $e=-\left|e\right|$. 
The collision integrals $\mathcal{I}_{R,L}$
can be obtained from the self-energies contributed by terms associated to the Feynman diagrams shown in Fig.~\ref{fig:sigma}. 
Each diagram represents multiple scattering processes, as the fermion internal lines carry species labels for 
chirality (right-, left-moving electrons) and Keldysh (retarded and advanced) sectors. 
Averaging over the Gaussian disorder potential, we recover the translational
symmetry and obtain the following collision integral for the right movers,
\begin{align}\label{IREval}
	\mathcal{I}_{R}\left(k\right)
	= 
	& W^{2}
	\intop_{k_{1},k_{3}}
	\left(k-k_{1}\right)^{2}\frac{2\Delta}{k_{3}^{2}+\Delta^{2}}\left\{ \begin{array}{l}
	\left[1-f_{R}\left(k\right)\right]\left[1-f_{R}\left(k_{1}\right)\right]f_{R}\left(k+k_{1}+k_{3}\right)f_{L}\left(k_{3}\right)\\
	-f_{R}\left(k\right)f_{R}\left(k_{1}\right)\left[1-f_{R}\left(k+k_{1}+k_{3}\right)\right]\left[1-f_{L}\left(k_{3}\right)\right]
	\end{array}\right\} 
\nonumber \\
	& -W^{2}
	\intop_{k_{1},k_{3}}
	k_{1}\left(k-2k_{1}-k_{3}\right)\frac{2\Delta}{k_{3}^{2}+\Delta^{2}}\left\{ \begin{array}{l}
	\left[1-f_{R}\left(k\right)\right]\left[1-f_{L}\left(k_{3}\right)\right]f_{R}\left(k_{1}\right)f_{R}\left(k-k_{1}-k_{3}\right)\\
	-f_{R}\left(k\right)f_{L}\left(k_{3}\right)\left[1-f_{R}\left(k_{1}\right)\right]\left[1-f_{R}\left(k-k_{1}-k_{3}\right)\right]
	\end{array}\right\} 
\nonumber \\
	& -W^{2}
	\intop_{k_{1},k_{3}}
	k_{3}\left(k_{1}-k_{3}\right)\frac{2\Delta}{k^{2}+\Delta^{2}}\left\{ \begin{array}{l}
	\left[1-f_{R}\left(k\right)\right]\left[1-f_{L}\left(k+k_{1}+k_{3}\right)\right]f_{L}\left(k_{1}\right)f_{L}\left(k_{3}\right)\\
	-f_{R}\left(k\right)f_{L}\left(k+k_{1}+k_{3}\right)\left[1-f_{L}\left(k_{1}\right)\right]\left[1-f_{L}\left(k_{3}\right)\right]
	\end{array}\right\} \,.
\end{align}
In this equation, $\intop_{k_{1},k_{3}} \equiv \int_{-\infty}^\infty (d k_1 / 2\pi) \int_{-\infty}^\infty (d k_3 / 2\pi)$, and 
we have introduced the disorder variance $\Delta \equiv 2\lambda^{2}/v_{F}^{2}$. 
The collision integral for the left movers can be obtained by the exchange of $f_{R}$ and $f_{L}$,
\begin{align}
	\mathcal{I}_{L}\left(k\right)
	= 
	& \mathcal{I}_{R}\left[f_{R}\leftrightarrow f_{L}\right]\,.
\end{align}

\subsection{Clean Limit}

In the clean limit $\Delta \rightarrow 0$, the momentum is conserved in each collision and 
Eq.~(\ref{IREval}) is simplified by replacing $2\Delta/(k^{2}+\Delta^{2}) \to 2\pi\delta\left(k\right)$.
Then we have 
\begin{align}
	\mathcal{I}_{R}\left(k\right)
	= 
	& 
	W^{2}
	\intop_{k_{1}}
	\left(k-k_{1}\right)^{2}
	\left\{ 
		\begin{array}{l}
			\left[1-f_{R}\left(k\right)\right]\left[1-f_{R}\left(k_{1}\right)\right]f_{R}\left(k+k_{1}\right)f_{L}\left(0\right)
		\\
			-
			f_{R}\left(k\right)f_{R}\left(k_{1}\right)\left[1-f_{R}\left(k+k_{1}\right)\right]\left[1-f_{L}\left(0\right)\right]
		\end{array}
	\right\} 
\nonumber \\
	& 
	-
	W^{2}
	\intop_{k_{1}}
	k_{1}\left(k-2k_{1}\right)
	\left\{ 
		\begin{array}{l}
			\left[1-f_{R}\left(k\right)\right]\left[1-f_{L}\left(0\right)\right]f_{R}\left(k_{1}\right)f_{R}\left(k-k_{1}\right)
			\\
			-
			f_{R}\left(k\right)f_{L}\left(0\right)\left[1-f_{R}\left(k_{1}\right)\right]\left[1-f_{R}\left(k-k_{1}\right)\right]
		\end{array}
	\right\} 
\nonumber \\
	& 
	-
	2\pi\delta\left(k\right)
	W^{2}
	\intop_{k_{1},k_{3}}
	k_{3}\left(k_{1}-k_{3}\right)
	\left\{ 
		\begin{array}{l}
			\left[1-f_{R}\left(0\right)\right]\left[1-f_{L}\left(k_{1}+k_{3}\right)\right]f_{L}\left(k_{1}\right)f_{L}\left(k_{3}\right)
			\\
			-
			f_{R}\left(0\right)f_{L}\left(k_{1}+k_{3}\right)\left[1-f_{L}\left(k_{1}\right)\right]\left[1-f_{L}\left(k_{3}\right)\right]
		\end{array}
	\right\} \,.
\end{align}
The 1PU scattering processes involve three particles of one chirality and another particle of the opposite chirality.
In the clean limit, the latter always resides \emph{exactly} at the Dirac point. This is the kinematics
depicted in Fig.~1(c) of the main text. As a result, the scattering is strongly suppressed when the chemical potential is 
tuned away from the edge Dirac point at low temperature. 

Integrating the background-subtracted kinetic equation (2) in the main text over momentum, 
we obtain the continuity equations for particle number,
\begin{subequations}\label{SM--nCont}
\begin{align}
	\partial_{t}n_{R}+v_{F}\partial_{x}n_{R} & =	\frac{eE}{2\pi}+I\,,
\\
	\partial_{t}n_{L}-v_{F}\partial_{x}n_{L} & =	-\frac{eE}{2\pi}-I\,.
\end{align}
\end{subequations}
The background-subtracted densities are defined via
\begin{align}\label{Densities}
	n_{R} 
	& =
	\intop_{k}
	\left[
		f_{R}\left(k\right)
		-
		\theta\left(-k\right)
	\right]
	=
	\frac{\mu_{R}}{2\pi v_{F}}\,,
\qquad 
	n_{L}
	=
	\intop_{k}
	\left[
		f_{L}\left(k\right)
		-
		\theta\left(k\right)
	\right]
	=
	\frac{\mu_{L}}{2\pi v_{F}}\,.
\end{align}
Here $\theta(k)$ denotes the Heaviside unit step function. 
In Eq.~(\ref{Densities}), we assume local equilibrium (ideal hydrodynamics \cite{SM--Landau}), 
so that 
$f_{R,L}(k) \equiv f(\tilde{\varepsilon}_{R,L},T_{R,L})$,
where $f$ is the Fermi-Dirac distribution [Eq.~(7)] and 
$\tilde{\varepsilon}_{R,L} = \pm v_F k - \mu_{R,L}$.
The imbalance force from 1-particle umklapp interaction is given by 
\begin{align}\label{SM--IDef}
	I 
	\equiv&\,
	\intop_{k} \mathcal{I}_{R}\left(k\right)
	=
	-\intop_{k}\mathcal{I}_{L}\left(k\right)
\nonumber\\
	=&\,
	\frac{\pi^2 W^2}{96 v_F^5}
	\left(f_L^0 - f_R^0\right)
	\left\{
		\begin{aligned}[r]
			\left[4 (v_F n_R)^2 + (k_B T_R)^2\right]
			\left[4 (v_F n_R)^2 + 9 (k_B T_R)^2\right]
		\\ 
			+
			\left[4 (v_F n_L)^2 + (k_B T_L)^2\right]
			\left[4 (v_F n_L)^2 + 9 (k_B T_L)^2\right]
		\end{aligned}
	\right\},
\end{align}
where $f^0_{R,L} \equiv \left[e^{- 2 \pi v_F n_{R,L} / (k_B T_{R,L})} + 1\right]^{-1}$.
Setting $v_F = k_B = 1$ recovers Eq.~(12) of the main text.

Multiplying the kinetic equation by $k$ and integrating over it,
we derive the continuity equations for momentum 
\begin{subequations}\label{SM--PCont}
\begin{align}
	\partial_{t}P_{R}+v_{F}\partial_{x}P_{R} & =e E \, n_{R},\\
	\partial_{t}P_{L}-v_{F}\partial_{x}P_{L} & =e E \, n_{L}.
\end{align}
\end{subequations}
Here 
\begin{equation}\label{PRPLEqs}
	P_{R}
	=
	\intop_{k}
	k \left[f_{R}\left(k\right) - \theta\left(-k\right)\right]
	=
	\frac{\pi k_{B}^{2}T_{R}^{2}}{12v_{F}^{2}}+\pi n_{R}^{2},
\qquad 
	P_{L}
	=
	\intop_{k}
	k \left[f_{L}\left(k\right) - \theta\left(k\right)\right]
	=
	-\frac{\pi k_{B}^{2}T_{L}^{2}}{12v_{F}^{2}}-\pi n_{L}^{2}.
\end{equation}
The friction forces 
$
	F_{in}^{R,L}=\intop_{k} k \, \mathcal{I}_{R,L}\left(k\right) = 0,
$
since the 1PU collisions separately conserve the momentum of right-
and left-moving electrons. 
Hence in the clean limit, the dissipation of the electric current does not come from the 
friction force on the electrons, but from the imbalance force $I$ [Eq.~(\ref{SM--nCont})].
This is a general conclusion for relativistic fermions
even if the friction force is nonzero. The friction force can dissipate
the momentum of the particles, but cannot change the velocity and thus
the current of the particles.


\subsection{Dirty Limit}

When there is disorder, the friction force does not vanish and we
have the continuity equations,
\begin{subequations}
\begin{gather}
	\partial_{t}n_{R}+v_{F}\partial_{x}n_{R} 
	= 
	\frac{e E}{2\pi}
	+
	I,
\qquad
	\partial_{t}n_{L}-v_{F}\partial_{x}n_{L}
	=
	-\frac{e E}{2\pi}
	-
	I,
\\
	\partial_{t}P_{R}+v_{F}\partial_{x}P_{R} 
	=
	e E \, n_{R}
	+
	F_{in}^{R},
\qquad
	\partial_{t}P_{L}-v_{F}\partial_{x}P_{L}
	=
	e E n_{L}
	+
	F_{in}^{L}.
\end{gather}
\label{eq:contidirty}
\end{subequations}

In the dirty limit, when the temperature is much lower than the 
characteristic disorder strength, $T\ll T_{\text{dis}} \equiv  v_{F} \Delta / k_{B}$,
we can approximate the Lorentzians in Eq.~(\ref{IREval}) as
\begin{equation}
	\frac{2 \Delta}{\Delta^{2}+k^{2}}\sim\frac{2}{\Delta}\,.
\end{equation}
Then we have the imbalance and friction forces,
\begin{align}\label{IDirty}
	I =&\,
	-
	\frac{\pi^{2}W^{2}}{360\Delta v_{F}^{5}}
	\left[
		\left(71T_{R}^{4}+50T_{R}^{2}T_{L}^{2}+71T_{L}^{4}\right)k_{B}^{4}\left(n_{R}-n_{L}\right)+120v_{F}^{2}k_{B}^{2}\left(T_{R}^{2}+T_{L}^{2}\right)\left(n_{R}-n_{L}\right)^{3}+48v_{F}^{4}\left(n_{R}-n_{L}\right)^{5}
	\right],
\\
\label{FinDirty}
	F_{in}^{R,L} 
	=&\, 
	-\frac{\pi^{3}W^{2}}{96\Delta v_F}
	\left\{ 
		\begin{array}{l}
			\frac{64}{5}
			\left(n_{R}+n_{L}\right) \left(n_{R}-n_{L}\right)^{5}
			+
			\frac{16}{3}
			\left(n_{R}-n_{L}\right)^{3}
			\left(\frac{k_B}{v_F}\right)^2		
			\left[
				\left(5n_{R}+7n_{L}\right)T_{R}^{2}
				+
				\left(7n_{R}+5n_{L}\right)T_{L}^{2}
			\right]
		\\
			+
			\frac{4}{15}
			\left(n_{R}-n_{L}\right)
			\left(\frac{k_{B}}{v_F}\right)^{4}	
			\left[
				\left(85 n_{R}+57 n_{L}\right)T_{R}^{4}
				+
				50
				\left(n_{R}+n_{L}\right)T_{R}^{2} T_{L}^{2}
				+
				\left(57 n_{R}+85 n_{L}\right)T_{L}^{4}
			\right]
		\\
			+
			\frac{1}{63}
			\left(\frac{k_{B}}{v_{F}}\right)^6
			\left[
				-199\left(T_{R}^{6}-T_{L}^{6}\right)
				+
				21T_{R}^{2}T_{L}^{2}\left(T_{R}^{2}-T_{L}^{2}\right)
			\right]
		\end{array}
	\right\}.
\end{align}

Since the system is isolated (no contact with a phonon bath, for example), the only violation 
of energy conservation is due to Joule heating,
\begin{align}\label{Joule}
	\partial_t u
	+
	\partial_x j_{\varepsilon}
	=
	j E,
\end{align}
where 
$u = v_F(P_R - P_L)$ is the internal energy density,
$j_{\varepsilon} = v_F^2(P_R + P_L)$ is the energy current, 
and 
$j = e v_F (n_R - n_L)$ is the electric current.

\section{Transport Coefficients}


\subsection{Summary: Ballistic to Hydrodynamic Crossover}
In this section, we discuss the thermoelectric transport \cite{Kuroda2008,Ghaemi2010,Blasi2020,Blasi2020a} 
of helical edge states via the chiral hydrodynamics.
We recover known results for the electrical conductivity, and obtain new ones for the thermoelectric power
and thermal conductance. These linear response results serve to benchmark the hydrodynamic formalism
employed to study nonequilibrium dynamics in the main text. Results are summarized here,
and derived in the subsequent subsections. 

We consider a setup wherein a helical edge is 
connected to electrically and/or thermally conducting leads.
The leads also serve as a thermal reservoir and equilibrate the fermions. 
The right movers originate from the left lead, held at temperature
$T_{1}$ and electrochemical potential $\bar{\mu}_{1}$.
Similarly, at the right end of the helical edge, 
the injected left movers have temperature $T_{2}$
and electrochemical potential $\bar{\mu}_{2}$. 
Here $\bar{\mu} \equiv \mu + e\phi(x)$ with $\phi$ the voltage;
the electrochemical potential is continuous between the leads and along the edge.

In electric transport, the two leads are held at the same temperature $T_{1,2} = T$;
in steady-state, Eq.~(\ref{SM--nCont}) reduces to
\begin{equation}\label{SM--muRLtoI}
	\partial_{x}\bar{\mu}_{R}
	=
	\partial_{x}\bar{\mu}_{L}
	=
	2\pi I.
\end{equation}
Here we have used the thermodynamic relation $\mu_{R,L} = 2 \pi v_F n_{R,L}$,
and
$I$ denotes the 1PU imbalance force. 
In the linear response regime
we obtain the conductance 
\begin{equation}\label{SM--GRes}
	G
	=
	(e^2/h)
	\left(
	1+l/l_{\text{el}}
	\right)^{-1},
\end{equation}
with $l$ the length of the helical edge and $l_{\text{el}}$ the
electric transport scattering length. In the clean limit, we find
\begin{equation}\label{SM--lelcRes}
	l_{\text{el}}^{\mathsf{c}}
	=
	\frac{
		96\pi v_{F}^{5}
		k_{B}T \cosh^{2}\left[{\mu}/{(2 k_B T)}\right]
	}{
		W^{2}
		\left(\mu^{2}+\pi^{2}k_{B}^{2}T^{2}\right)\left(\mu^{2}+9\pi^{2}k_{B}^{2}T^{2}
	\right)}.
\end{equation}
On the other hand, the scattering length of electric transport in the dirty limit 
\cite{Fiete2006}
is given by 
\begin{equation}\label{SM--leldRes}
	l_{\text{el}}^\mathsf{d}
	=
	\left(
		15 \Delta v_{F}^{6}
	\right)
	/
	\left(
		8\pi^{2}W^{2}k_{B}^{4}T^{4}
	\right),
\end{equation}
valid for $k_B T \ll v_F \Delta$.
When the size of sample $l \ll l_{\text{el}}$, 
the transport is nearly ballistic.
Ohmic conduction is recovered in the opposite limit $l \gg l_{\text{el}}$,
where the results agree with previous work \cite{Kainaris2014,Chou2015}.

In a thermal conductivity measurement, the two leads are held at
the same electrochemical potential, but different temperatures,
and no electric current is allowed to flow. 
Solving the relation between the thermal current and the temperature difference between 
the two leads, we obtain the thermal conductance 
\begin{equation}\label{SM--KRes}
	\mathsf{K}
	=
	\left[
		\pi^{2}k_{B}^{2}T
		/
		(3 h)
	\right]
	\left(
		1+l/l_{\text{th}}
	\right)^{-1},
\end{equation}
where $l_{\text{th}}$ is the scattering length for thermal transport.
In the clean limit 
$
	l_{\text{th}}^\mathsf{c}
	=
	(\pi^2/3)
	\left(k_{B} T / \mu \right)^2
	l_{\text{el}}^\mathsf{c},
$
while in the dirty limit 
$
	l_\text{th}^\mathsf{d}
	=
	(7/15)
	l_{\text{el}}^\mathsf{d}.
$

The Wiedemann-Franz law is recovered for $l\ll l_{\text{el},\text{th}}$,
where
$\mathsf{K}/(G T) = (\pi^2/3)\left(k_B / e\right)^2 \equiv L$ is the Lorenz number. 
However, the Wiedemann-Franz law is violated for longer devices, since electric and 
thermal transport are associated to different scattering lengths. 
For long edges $l\gg l_{\text{el,th}}$, we have 
\begin{equation}\label{SM--LRatioRes}
	\frac{\mathsf{K}_{\text{clean}}}{G_{\text{clean}}T}=\frac{\pi^2k_{B}^{2}T^{2}}{3\mu^{2}}L\,,\qquad\frac{\mathsf{K}_{\text{dirty}}}{G_{\text{dirty}}T}=\frac{7}{15}L\,.
\end{equation}
For a clean helical liquid, the thermal conductance $\mathsf{K}$ is proportional
to $G \, T^{3}$ rather than $G \, T$.
In the dirty limit, $\mathsf{K}$ is proportional to $G \, T$, but
with the ratio $(7/15) L$. 

In a thermoelectric power measurement, a voltage is generated by
the temperature difference applied across the two leads when no electric current flows
in the sample. In the clean limit, we find the Seebeck coefficient 
\begin{equation}\label{SM--SRes}
	S
	=
	-
	\Delta V / \Delta T
	=
	\left(
		s / \rho
	\right)
	\left(
		1 + l_{\text{th}}^\mathsf{c}/l
	\right)^{-1}.
\end{equation}
Here $s / \rho = (\pi^2 k_B^2 T)/(3 e \mu)$ is the ratio of the
equilibrium entropy density $s$ to the charge density $\rho = e (n_R + n_L)$.
The Seebeck coefficient goes to zero as $l / l_{\text{th}}^\mathsf{c}$ in the short device $l \rightarrow 0$ 
limit.  On the other hand, the Seebeck coefficient always vanishes in the dirty limit.

\subsection{Clean Limit}

\subsubsection{Electric Transport}

In an electric transport measurement, two leads with temperature $T$ and electrochemical potentials $\bar{\mu}_{1}$ and $\bar{\mu}_{2}$
are connected to the sample. Then we obtain Eq.~(\ref{SM--muRLtoI}) for the steady-state, which reads
[via Eq.~(\ref{SM--IDef})]
\begin{equation}
	\partial_{x}\bar{\mu}_{R}
	=
	\partial_{x}\bar{\mu}_{L}
	=
	2\pi I
	\simeq
	-
	\frac{
		W^{2}\sinh\left(\frac{\beta\delta\mu}{2}\right) \left(\mu^{2}+\pi^{2}k_{B}^{2}T^{2}\right)\left(\mu^{2}+9\pi^{2}k_{B}^{2}T^{2}\right)
	}{
		48\pi v_{F}^{5} \cosh^{2}\left(\frac{\beta\mu}{2}\right)
	},
\end{equation}
where $\beta = (k_B T)^{-1}$, and 
$\bar{\mu}_{R,L}(x) \equiv \mu_{R,L}(x) + e\phi(x)$ denote the electrochemical
potentials. Here $\phi(x)$ is the electric potential from the gate, bias voltage, and 
internal inhomogeneity. We also assume that
the chemical potentials of the right and left movers are close to each other: 
$
	\delta\mu \equiv \mu_{R} - \mu_{L}
	\ll
	\left|\mu\right|,
$ 
where
$\mu \equiv (\mu_{R} + \mu_L)/2$. 

Assuming slow spatial variation along the edge and using the boundary
conditions 
$\bar{\mu}_{R}(0) = \bar{\mu}_{1}$
and
$\bar{\mu}_{L}(l) = \bar{\mu}_{2}$,
where $l$ is the length of the edge, we have
\begin{align}
	\bar{\mu}_{R}\left(x\right) 
	=
	-
	2 k_B T
	\sinh\left(\pi j_n \beta\right)
	\left(
	\frac{x}{l_{\text{el}}^\mathsf{c}}
	\right)
	+
	\bar{\mu}_{1},
\qquad
	\bar{\mu}_{L}\left(x\right) 
	=
	-
	2 k_B T
	\sinh\left(\pi j_n \beta\right)
	\left(
	\frac{x - l}{l_{\text{el}}^\mathsf{c}}
	\right)
	+
	\bar{\mu}_{2}.
\end{align}
Here $j_n=\delta\mu/(2\pi)$ is the particle number current,
and $l_{\text{el}}^\mathsf{c}$ is the clean electric scattering length
[Eq.~(\ref{SM--lelcRes})].
Then we have 
\begin{equation}
	2 \pi j_n
	=
	-
	\left(
	\frac{l}{l_{\text{el}}^\mathsf{c}}
	\right)
	2 k_B T
	\sinh\left(\pi j_n \beta\right)
	+
	e \Delta V,
\label{eq:jn}
\end{equation}
where $\Delta V = (\bar{\mu}_1 - \bar{\mu}_2)/e$ is the voltage drop across 
the leads. 
In the linear response regime with $|j_{n}| \ll k_{B}T$,
we obtain the electric current, 
\begin{equation}
	j 
	= 
	e
	j_{n}
	=
	\frac{e^{2}}{h}
	\left(\frac{\Delta V}{1+l/l_{\text{el}}^\mathsf{c}}\right),
	\label{eq:j}
\end{equation}
where we have restored Planck's constant. 
This determines the electrical conductance in Eq.~(\ref{SM--GRes}).

\subsubsection{TEP}

In a measurement of the thermoelectric power (TEP), the electrical current is exactly zero.
Therefore $n_R = n_L$, so that $\mu_R = \mu_L \equiv \mu$. 
The hydrodynamic equations (\ref{SM--PCont}) and (\ref{SM--nCont}) then imply that 
\begin{align}\label{TEP--1}
	\frac{\pi^2 k_{B}^{2}\partial_{x} \left(T_{R}^{2}\right)}{6}  
	=
	\frac{\pi^2 k_{B}^{2}\partial_{x} \left(T_{L}^{2}\right)}{6}
	=
	-
	\mu \partial_{x}\bar{\mu} 
	=
	-
	2 \pi \mu I,
\end{align}
where $\partial_x \bar{\mu} \equiv \partial_x \mu - e E$. 
Therefore in the linear response regime, we have
\begin{align}\label{TEP--2}
\begin{aligned}
	\partial_x\left(T_R^2 - T_L^2\right) 
	=&\, 
	0,
\\
	\partial_x\left(T_R^2 + T_L^2\right) 
	=&\, 
	-
	\frac{24 \mu}{\pi k_B^2} I
\\
	\simeq&\, 
	-
	\frac{4}{l_{\text{th}}^\mathsf{c}}
	T
	\left(T_R - T_L\right),
\end{aligned}
\end{align}
where we have linearized the imbalance force $I$ [Eq.~(\ref{SM--IDef})] in $(T_R - T_L)$,
and where the thermal scattering length $l_{\text{th}}^\mathsf{c}$ is 
\begin{align}
	l_{\text{th}}^\mathsf{c}
	=
	\left(\frac{\pi k_B T}{\mu}\right)^2
	\frac{l_{\text{el}}^\mathsf{c}}{3}.
\end{align}
Therefore
\begin{subequations}\label{TRTL_Profile}
\begin{align}
	T_R(x) 
	=&\,
	\frac{1}{l_{\text{th}}^\mathsf{c}}
	\left(T_L - T_R\right)(0)
	\,
	x
	+
	T_R(0),
\\
	T_L(x) 
	=&\,
	\frac{1}{l_{\text{th}}^\mathsf{c}}
	\left(T_L - T_R\right)(0)
	\,
	(x - l)
	+
	T_L(l),
\end{align}
\end{subequations}
where the left and right leads appear at $x = 0$ and $x = l$, respectively. 

Eqs.~(\ref{TEP--1}) and (\ref{TEP--2}) then imply that 
\begin{align}
	\bar{\mu}_1 
	-
	\bar{\mu}_2
	=
	e 
	\Delta V
	=
	\frac{\pi^2 k_B^2 T}{3 \mu}
	\left(\frac{l}{l_{\text{th}}^\mathsf{c}}\right)
	\left(T_L - T_R\right)(0),
\end{align}
where $\Delta V$ is the voltage drop across the leads. 
We can eliminate $T_L(0)$ using Eq.~(\ref{TRTL_Profile}),
\begin{align}\label{TL0Elim}
	T_L(0) 
	=&\,
	\left(
	1
	+
	\frac{l}{l_{\text{th}}^\mathsf{c}}
	\right)^{-1}
	\left[
	\left(\frac{l}{l_{\text{th}}^\mathsf{c}}\right)
	T_R(0)
	+
	T_L(l)
	\right],
\end{align}
so that finally
\begin{align}
	e 
	\Delta V
	=
	-
	\frac{\pi^2 k_B^2 T}{3 \mu}
	\left(
	1
	+
	\frac{l_{\text{th}}^\mathsf{c}}{l}
	\right)^{-1}
	\Delta T,
\end{align}
where the temperature difference between the leads is 
$\Delta T \equiv T_R(0) - T_L(l)$. 
The TEP is determined by the Seebeck coefficient
\begin{align}
	S
	=
	-
	\frac{\Delta V}{\Delta T}
	=
	\frac{\pi^2 k_B^2 T}{3 \mu e}
	\left(
	\frac{l}{
	l
	+
	l_{\text{th}}^\mathsf{c}
	}
	\right),
\end{align}
which is Eq.~(\ref{SM--SRes}). 
The Seebeck coefficient vanishes in the limit $l/l_{\text{th}}^\mathsf{c} \rightarrow 0$
(a very short wire, or zero umklapp interaction strength), 
since there is no thermopower without irrelevant elastic scattering \cite{SM--Kane1996}.

\subsubsection{Thermal Conductance}

Using the first law of thermodynamics, 
\begin{equation}
	d U = T dS +\mu_R \, d N_R + \mu_L \, d N_L,
\end{equation}
we define the thermal current,
\begin{align}
	j_{q} 
	\equiv&\,
	j_{\varepsilon}
	-
	\mu_R \, j_{R}
	-
	\mu_L \, j_{L}
	=
	v_{F}^{2}\left(P_{R}+P_{L}\right)
	-
	\mu_{R} \, (v_F n_R)
	-
	\mu_{L} \, (-v_F n_L)
\nonumber\\
	=&\,
	\frac{\pi k_{B}^{2}\left(T_{R}^{2}-T_{L}^{2}\right)}{12}
	-
	\pi\left(n_{R}^{2}-n_{L}^{2}\right)v_{F}^{2},
\end{align}
where $j_\varepsilon$ is the energy current [Eq.~(\ref{Joule})]. 

In a thermal conductance measurement, the electric current is zero so that $n_{R} = n_{L}$. 
Then we have 
\begin{equation}
	j_{q}
	=
	\frac{\pi k_{B}^{2}\left(T_{R}^{2}-T_{L}^{2}\right)}{12}
	\simeq\frac{\pi k_{B}^{2}T\left(T_{R}-T_{L}\right)}{6},
\end{equation}
assuming linear response. 
Eq.~(\ref{TL0Elim}) then implies that 
\begin{align}
	\left(T_{R}-T_{L}\right)
	=
	\left(1 + \frac{l}{l_{\text{th}}^\mathsf{c}}\right)^{-1}
	\Delta T,
\end{align}
so that
\begin{equation}
	j_{q}
	=
	\frac{\pi^2 k_{B}^{2}T}{3 h}
	\left(
		\frac{l_{\text{th}}^\mathsf{c}}{l + l_{\text{th}}^\mathsf{c}}
	\right)
	\Delta T,
\end{equation}
where we have restored Planck's constant. This gives the thermal conductance 
in Eq.~(\ref{SM--KRes}).


\subsection{Dirty Limit}

\subsubsection{Electric Transport}

In the dirty limit $k_B T \ll v_F \Delta$, the imbalance force in electric transport ($T_R = T_L = T$) is expressed as
\begin{equation}
	I
	=
	-\frac{2\pi^{2}W^{2}\left(n_{R}-n_{L}\right)}{15\Delta v_{F}^{5}}\left[4k_{B}^{4}T^{4}+5v_{F}^{2}k_{B}^{2}T^{2}\left(n_{R}-n_{L}\right)^{2}+v_{F}^{4}\left(n_{R}-n_{L}\right)^{4}\right]\,.
  \label{eq:Idirty}
\end{equation}
Then we have the steady state equation for the electrochemical potentials,
\begin{equation}
  \partial_{x}\bar{\mu}_{R,L}=-\frac{4\pi^{3}W^{2}\left(n_{R}-n_{L}\right)}{15\Delta v_{F}^{5}}\left[4k_{B}^{4}T^{4}+5v_{F}^{2}k_{B}^{2}T^{2}\left(n_{R}-n_{L}\right)^{2}+v_{F}^{4}\left(n_{R}-n_{L}\right)^{4}\right]\,.
\end{equation}
 Solving the differential equations, we have
 \begin{equation}
   2\pi j_n\left[ 1+\frac{l}{l_\text{el}^\mathsf{d}}\left( 1+\frac{5j_n^2}{4k_B^2T^2}+\frac{j_n^4}{4k_B^4T^4} \right) \right] = e\Delta V\,.
   \label{eq:jndirty}
 \end{equation}
 Here the electric transport scattering length in dirty limit is given by
 \begin{equation}
   l_{\text{el}}^\mathsf{d}=\frac{15\Delta v_{F}^{6}}{8\pi^{2}W^{2}k_{B}^{4}T^{4}}\,.
   \label{eq:leld}
 \end{equation}
In the linear response regime $\left|j_{n}\right|\ll k_{B}T$, the electric current is given by
\begin{equation}
  j=\frac{e^{2}}{h}\left(\frac{\Delta V}{1+l/l_{\text{el}}^\mathsf{d}}\right),
\end{equation}
with Planck's constant restored. This gives the electric conductance in 
Eqs.~(\ref{SM--GRes}) and (\ref{SM--leldRes}).

\subsubsection{TEP}

The total particle current vanishes in thermal transport, i.e. $n_{R}-n_{L}=0$, which implies
vanishing imbalance force $I=0$ in the dirty limit [Eq.~(\ref{IDirty})]. In steady state, we have
\begin{equation}\label{IZeroDirty}
	\partial_{x}\bar{\mu}_{R,L}=2\pi I=0\,,
\end{equation}
which holds for arbitrary temperatures for the right and left movers. 
Thus the thermoelectric power vanishes in the dirty limit, where the disorder potential scattering 
dominates the irrelevant inelastic scattering.

\subsubsection{Thermal Conductance}

With $n_R=n_L$ in thermal transport,  the friction force is obtained as [Eq.~(\ref{FinDirty})]
\begin{equation}
  F_{in}^{R,L}  =\frac{\pi^{3}W^{2}k_B^6}{6048 v_{F}^{7}\Delta}\left[199\left(T_R^6-T_L^6\right)-21T_R^2T_L^2\left(T_R^2-T_L^2\right)\right]\,.
  \label{eq:Findirty}
\end{equation}
Combining Eqs.~\eqref{eq:contidirty}, \eqref{IZeroDirty} and \eqref{eq:Findirty}, we obtain the 
steady state equation of temperatures,
\begin{equation}
  \partial_{x}T_{R}^{2}=\partial_{x}T_{L}^{2}=-\frac{\pi^{2}W^{2}k_{B}^{4}}{504 v_{F}^{6}\Delta}\left(199T_{R}^{4}+178T_{R}^{2}T_{L}^{2}+199T_{L}^{4}\right)\left(T_{R}^{2}-T_{L}^{2}\right)\,.\label{eq:de_TRL}
\end{equation}
Similar to the derivation in the clean limit, we have
\begin{equation}
  T_{R}^{2}-T_{L}^{2}=\frac{T_{R}^{2}(0)-T_{L}^{2}(l)}{1+l/l_{\text{th}}^\mathsf{d}}\,,
\end{equation}
where $l_{\text{th}}^\mathsf{d}$ is the thermal transport scattering length,
\begin{equation}
	l_{\text{th}}^\mathsf{d}
	=
	\frac{7\Delta v_{F}^{6}}{8\pi^{2}W^{2}k_{B}^{4}T^{4}} 
	= 
	\frac{7}{15}l_\text{el}^\mathsf{d}\,.
	\label{eq:lthd}
\end{equation}
The thermal current is then obtained as
\begin{equation}
	j_{q}
	\simeq
	\frac{\pi^{2}k_{B}^{2}T}{3h}\left(\frac{\Delta T}{1+l/l_{\text{th}}^\mathsf{d}}\right),
\end{equation}
with Planck's constant restored. 
This gives the dirty limit thermal conductance in 
Eqs.~(\ref{SM--KRes}) and (\ref{SM--LRatioRes}).

\section{Shock Wave Dynamics}

\subsection{Setup \label{Sec:Setup}}

An initial strong electric field pulse along the helical edge induces an imbalance of left and right movers 
via the chiral anomaly. We assume that the system is initially in equilibrium, 
doped to the edge Dirac point ($n_R=n_L=0$), with some low but nonzero temperature $T_R=T_L=T_0 > 0$.

Ignoring the imbalance relaxation, which is reasonable for short, high-intensity pulse, 
the chiral density response to the field $E(t,x)$ is given by
\begin{equation}
	n_{R,L}(t,x) 
	= 
	\pm 	
	\frac{1}{2\pi}\int_0^t dt' eE[t',x\mp v_F(t-t')].
	\label{eq:nRL}
\end{equation}
We consider a Gaussian electric pulse in space and time
\begin{equation}
	eE(t,x) 
	= 
	\frac{2\sqrt{\pi} \Delta S}{\tau\xi}  
	\exp{\left[ -\frac{x^2}{\xi^2}-\frac{(t-\tau)^2}{\tau^2} \right]}\,,
  \label{eq:eE}
\end{equation}
where $\Delta S$ is the action of the force. 
For $t\gg \tau$, we have 
\begin{equation}
	n_{R,L}(t,x)
	\simeq
	\pm
	\left(\frac{\Delta S}{\xi_\tau}\right) 
	\exp{\left\{ -\frac{\left[ x\mp v_F(t-\tau) \right]^2}{\xi_\tau^2} \right\}}\,,
  \label{eq:nRLtx1}
\end{equation}
where $\xi_\tau \equiv \sqrt{\xi^2 + v_F^2\tau^2}$. 
For a pulse of duration $\tau\sim 1$ ps and the edge Fermi velocity $v_F \sim 10^5 \text{m}/\text{s}$, 
we have $v_F \tau \sim 0.1$ $\mu\text{m}$, which is comparable to a natural edge segment length $\xi$ that could be
selectively exposed to THz radiation. The action $\Delta S \propto \max(e V) \times \tau$, the 
potential peak amplitude times the pulse duration. 

In the absence of relaxation, the temperatures of the right and left movers are not changed by the 
external electric field. The collisionless equations of motion read
\begin{subequations}\label{eq:EOM_nocol}
\begin{align}
	(\partial_t + v_F\partial_x)n_R &= \frac{eE}{2\pi}, 
\\
	(\partial_t +v_F\partial_x)P_R &= e E \, n_R = \pi(\partial_t + v_F\partial_x)n_R^2,
\end{align}
\end{subequations}
so that the change $\Delta P_R = \pi \Delta n_R^2$. 
Since $P_R = \pi(k_B^2 T_R^2 / 12 v_F^2 + n_R^2)$
[Eq.~(\ref{PRPLEqs})], we must have $\Delta T_R^2 = 0$. 
Similarly $\Delta T_L^2 =0$. 
We assume that interactions are sufficiently weak so as to allow right- and left-propagating packets to separate. 
Focusing on the right-moving packet, we have the effective initial conditions
\begin{equation}
	n_R(0,x) 
	\equiv 
	n_{R}^{\text{\tiny (0)}}(x)
	=
	n_0 \, \exp{\left(-x^2/\xi^2\right)},
\qquad 
	n_L(0,x)
	\equiv
	n_{L}^{\text{\tiny (0)}}(x)
	= 0,
\label{eq:nR0x}
\end{equation}
and $T_R = T_L = T_0$. Here we have shifted the $x$-origin to the center of the right-moving packet.

The subsequent evolution in the absence of electric field incorporates the imbalance force,
\begin{subequations}
\begin{align}
  (\partial_t \pm v_F \partial_x) n_{R,L} &= \pm I[t,x,n_R,n_L,T_R,T_L],
\label{eq:EOMnR}
\\
  (\partial_t \pm v_F \partial_x) P_{R,L} &= 0.
\end{align}
\end{subequations}
We set $v_F = k_B = 1$. The conserved momenta are expressed in terms of the initial densities 
and temperatures,
\begin{subequations}
\begin{align}
	P_R(t,x) 
	=&\, 
	\pi\left\{ 
		\frac{T_0^2}{12} + \left[ n_R^{\text{\tiny (0)}}(x - t) \right]^2 \right\},
\\
	P_L(t,x) 
	=&\, 
	-\
	\pi \frac{T_0^2}{12}.
  \label{eq:PRLtx}
\end{align}
\end{subequations}
Using Eq.~(\ref{PRPLEqs}), the temperature $T_{R,L}$ can also be expressed as
\begin{subequations}\label{eq:TRLtx}
\begin{align}
	T_R(t,x) 
	=&\, 
	\left[T_0^2 + 12 \left\{ \left[ n_R^{\text{\tiny (0)}}(x - t) \right]^2 - \left[ n_R(t,x) \right]^2 \right\} \right]^{1/2},
\\
	T_L(t,x) 
	=&\, 
	\left[ T_0^2 -12  \left[ n_L(t,x) \right]^2 \right]^{1/2}.
\end{align}
\end{subequations}
The problem reduces to solving the coupled density equations (\ref{eq:EOMnR}), using
the 1PU imbalance force in the clean limit [Eq.~(12) in the main text], 
with the temperature profiles determined by Eq.~(\ref{eq:TRLtx}).


\subsection{Quantum Quench of a Topological Insulator under a Tilted Potential \label{sec:SM--Quench}}

The charge packet formation described in Sec.~\ref{Sec:Setup} relies only upon the
chiral anomaly for the continuum theory of a single edge, 
encoded in the term $\pm E/2\pi$ on the right-hand side of Eq.~(3) in the main text.
For a helical edge, the formation of a local right-left density imbalance is equivalent to a 
local spin accumulation. In the absence of Rashba spin-orbit coupling (RSOC), this must be compensated
by an equal and opposite spin accumulation elsewhere. 

Here we consider the application of a spatially homogeneous electric field pulse across a 2D topological insulator
sample, using the Kane-Mele lattice model \cite{SM--Kane2005a}. We numerically solve for the evolution of the charge and spin densities induced
by quenching on a transient electric potential gradient. We show that for weak or vanishing RSOC
and for a peak electric field strength that is not too large (compared to the bulk gap), the dynamics are 
entirely consistent with the chiral anomaly. In this case, equal and opposite spin accumulations appear
along opposite edges that are aligned with the external field. This leads to four
isolated chiral charge packets as shown in Fig.~1(b) of the main text. 
In what follows, we ignore interactions [the 1PU inelastic scattering encoded by $I$ in Eq.~(3) of the main text], 
which are responsible for shock formation.

\begin{figure}[t]
\centering
\includegraphics[width=0.7\textwidth]{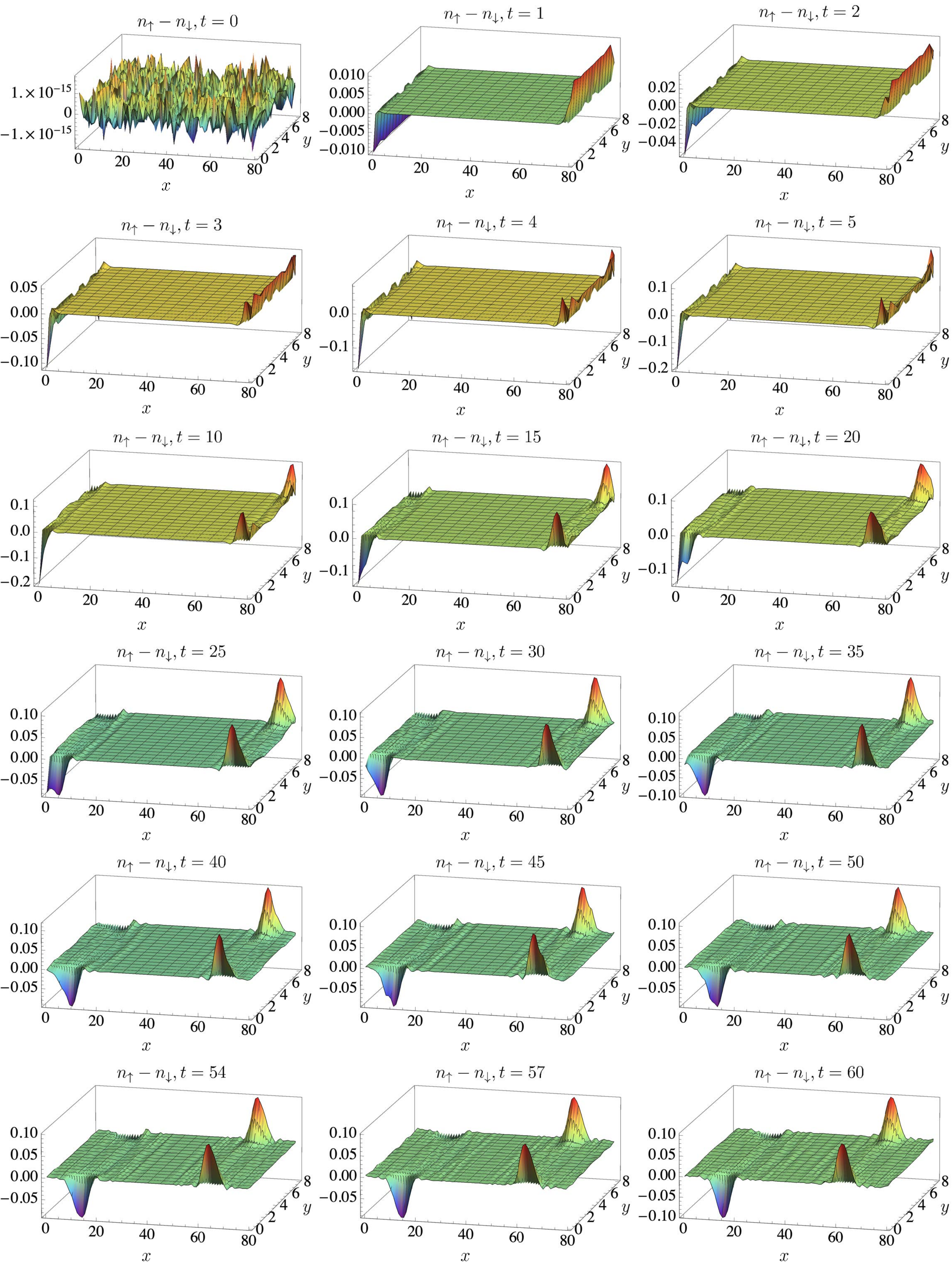}
\caption{The spin density in the lattice quench dynamics for the Kane-Mele model subjected
to a transient, homogeneous electric field, as described above Eq.~(\ref{eq:Vyt}).
The electric field is applied along the short y-direction of the lattice, leading to a positive
(negative) spin accumulation at the right (left) edge. 
Here the Kane-Mele model parameters [Eq.~(\ref{eq:HKM})] are $\lambda_\text{SO}=0.06$ and $\lambda_R=0$ 
(so that the total $z$-spin is conserved), 
while the electric potential gradient in Eq.~(\ref{eq:Vyt}) has $E_0=0.2$ and $t_0=3$.}
\label{fig:spin}
\end{figure}

\begin{figure}[t]
\centering
\includegraphics[width=0.7\textwidth]{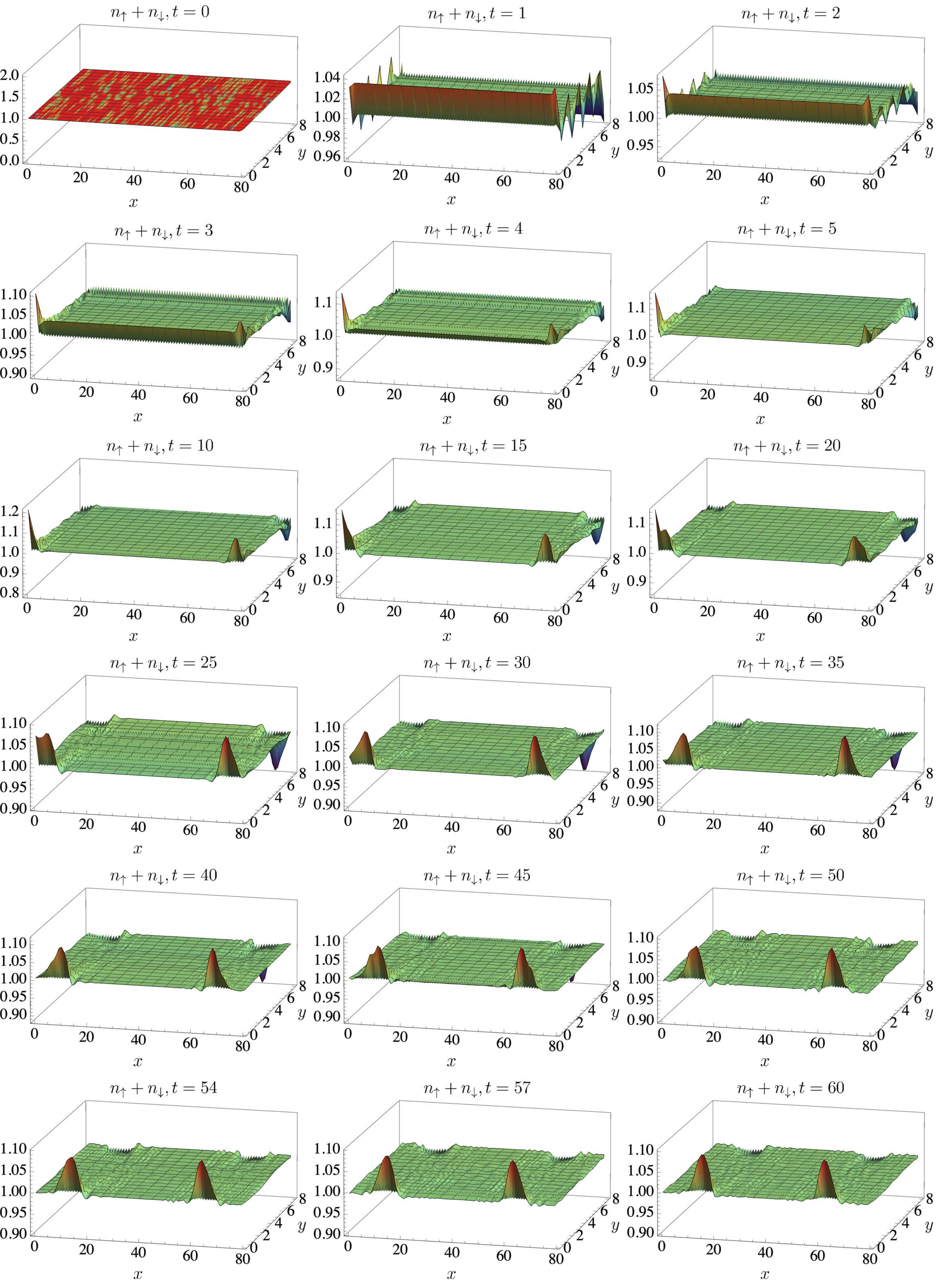}
\caption{The number density in the quench dynamics with $\lambda_\text{SO}=0.06$, $\lambda_R=0$, $E_0=0.2$ and $t_0=3$.}
\label{fig:density}
\end{figure}

The Kane-Mele model on honeycomb lattice is given by \cite{SM--Kane2005a}
\begin{equation}
  H_\text{KM} = \sum_{\langle ij\rangle}c^\dagger_{i\alpha} c_{j\alpha} + i\lambda_\text{SO}\sum_{\langle\langle ij\rangle\rangle}\nu_{ij} c^\dagger_{i\alpha}s_z^{\alpha\beta}c_{j\beta} + i\lambda_R\sum_{\langle ij\rangle}c^\dagger_{i\alpha}\left( \mathbf{s}\times\hat{\mathbf{d}}_{ij} \right)_z^{\alpha\beta}c_{j\beta}\,,
  \label{eq:HKM}
\end{equation}
with nearest-neighbor (NN) hopping set to $1$, 
$\lambda_\text{SO}$ is the next-nearest-neighbor (NNN) spin-orbit coupling, 
and $\lambda_R$ is the Rashba spin-orbit coupling. 
Here 
$\nu_{ij} = \frac{2}{\sqrt{3}}\left( \hat{\mathbf{d}}_1\times\hat{\mathbf{d}}_2 \right)_z=\pm 1$, 
with $\hat{\mathbf{d}}_1$ and $\hat{\mathbf{d}}_2$ the two unit NN bond vectors connecting NNN sites $i$ and $j$, 
and $\hat{\mathbf{d}}_{ij}$ denotes the unit vector connecting NN sites $i$ and $j$. 
In Eq.~(\ref{eq:HKM}), $s_i$ ($i=x,y,z$) denote the Pauli matrices acting on physical spin, 
and the Einstein summation convention is used for the spin indices $\alpha$ and $\beta$. 
The Kane-Mele model hosts topologically protected edge modes with nonzero $\lambda_\text{SO}$. 
The Rashba term $\lambda_R$ mixes the Haldane models in the spin-up and -down sectors and does not break the nontrivial topology of the system,
so long as it is not too large \cite{SM--Kane2005a}.

Now we consider the quantum quench dynamics of the Kane-Mele model on a rectangular slab of the honeycomb lattice 
with $N_s=2\times80\times 10$ sites. The initial state is the half-filling state with all single electron states below the Dirac point filled,
\begin{equation}
  \Psi_0 = \psi_1\otimes \psi_2\otimes\cdots\otimes\psi_{N_H/2}\,,
  \label{eq:Psi0}
\end{equation}
with $N_H = 2N_s$ being the dimension of the Hilbert space.
An electric field is then applied across the sample, polarized along the short $y$-direction of the rectangular slab
[see Fig.~1(a) in the main text]. The field is encoded in a tilted potential, which subsequently decays in time:
\begin{equation}
	V(t) = E_0 \, y \, e^{-t/t_0}.
	\label{eq:Vyt}
\end{equation}
The system evolves under the time-dependent Schr\"{o}dinger equation,
\begin{equation}
  i\frac{\partial \Psi(t)}{\partial t} = H(t) \Psi(t)\,,
  \label{eq:Schro}
\end{equation}
with $H(t) = H_\text{KM} + V(t)$. 

The quantum quench dynamics are computed using the 4th-order Runge-Kutta method
with $\lambda_\text{SO}=0.06$, $E_0=0.2$ and $t_0=3$. 
The titled potential is weak and cannot excite bulk states.

In Fig.~\ref{fig:spin} and \ref{fig:density}, we show the spin and number density of the system in the 
quench dynamics with vanishing Rashba term $\lambda_R=0$. 
The uniform electric field applied along the short $y$-direction of the rectangular slab
induces equal and opposite spin accumulations along the short left and right edges. 
On the right edge, the electric field increases the density of the spin-up electrons (moving clockwise) 
and decreases that of spin-down electrons (moving counterclockwise). 
During the transient application of the field, the total number density increases (decreases)
along the long upper (long lower) edge due to the tilted potential, but maintains its initial value in the bulk.
This indicates that the bulk states do not significantly participate in the dynamics due to the bulk energy gap, 
which is much greater than the potential differences between neighboring sites. 
During the transient application of the field, 
some oscillating behavior emerges in the spin and number density on the short right and left edges, 
due to the spatial variation of edge states. 

After the titled potential is turned off, the spin-up and -down electrons in the bumps split into chiral components, and turn 
around the corners. 
The induced bump on the left edge has the same number density and opposite spin density compared to that of right edge, 
consistent with the chiral anomaly.
The separated spin-up and -down bumps have a smooth shape, with a packet width close to the length of the 
short edge of the sample (the length of the short edge is much smaller than that of the long edge). 
After rounding the corners, the spin-up and -down bumps propagate ballistically with speed given by the Fermi velocity of the edge states.

The Rashba term $\lambda_R$ breaks the $S^z$ symmetry, and asymmetric features appear for the spin-up and -down bumps induced by the chiral 
anomaly. However, the qualitative features of the quench dynamics with weak Rashba coupling
(not depicted here) remain the same as the case without RSOC.

The 1PU interaction is not considered in these quench dynamics, but the dynamics remains unchanged as long as the scattering length of the 
1PU interaction is much larger than the length of the short edge. Considering the 1PU interaction, the subsequent charge-packet evolution 
along the long edges is then captured by the hydrodynamic equations, and we can consider the right- and left-moving bumps separately.


\subsection{Dynamics from Numerical Solution of the Hydrodynamic Equations}

The hydrodynamic equations can be written in momentum space,
\begin{subequations}
\label{eq:dnk}
\begin{align}
	\dot{n}_R(k) + i k n_R(k) =&\, I(k), 
\\
	\dot{n}_L(k) - i k n_L(k) =&\, -I(k).
\end{align}
\end{subequations}
We consider a system of finite length $L$, with periodic boundary conditions;
thus momenta are therefore quantized, $k_n = 2 \pi n / L$, with 
$n \in \{-M/2 + 1,-M/2 + 2,\ldots,M/2\}$. 
We choose 
$L = 200 \xi$, 
where $\xi$ is the size of the initial density wave [Eq.~(\ref{eq:nR0x})],
and $M = 20000$.  
This corresponds to a real space discretization of 
$\Delta x \equiv L/M = 0.01 \xi$.  

Moreover, we can improve the real space resolution by incorporating smaller $\Delta x$, while maintaining the same momentum cutoff 
$k_\mathsf{cutoff}= 10000 \pi / L$ to avoid a numerical instability at 
$\max{\left(\left|k\right|\right)} \Delta t > 1$, with $\Delta t$ being the time step used in the numerics.
This does not change the accuracy of the numerical method, since the dynamics is determined by the slow-varying modes in momentum space, 
but allows us to evaluate quantities like the speed of the shock front relative to the right-moving lightcone, 
where better resolution in position $x$ is needed.

We compute the dynamics of the right-moving excess density packet by solving the coupled ODEs in Eq.~(\ref{eq:dnk}) 
via the 4th-order Runge-Kutta method. 
The imbalance force $I(k)$ is obtained as the Fourier transform of the real space expression $I(x)$
[Eq.~(12) in the main text], which is a function of $n_{R,L}(x)$.
The latter are calculated as the inverse Fourier transforms of $n_{R,L}(k)$.

\begin{figure}[b!]
\centering
\includegraphics[width=0.8\textwidth]{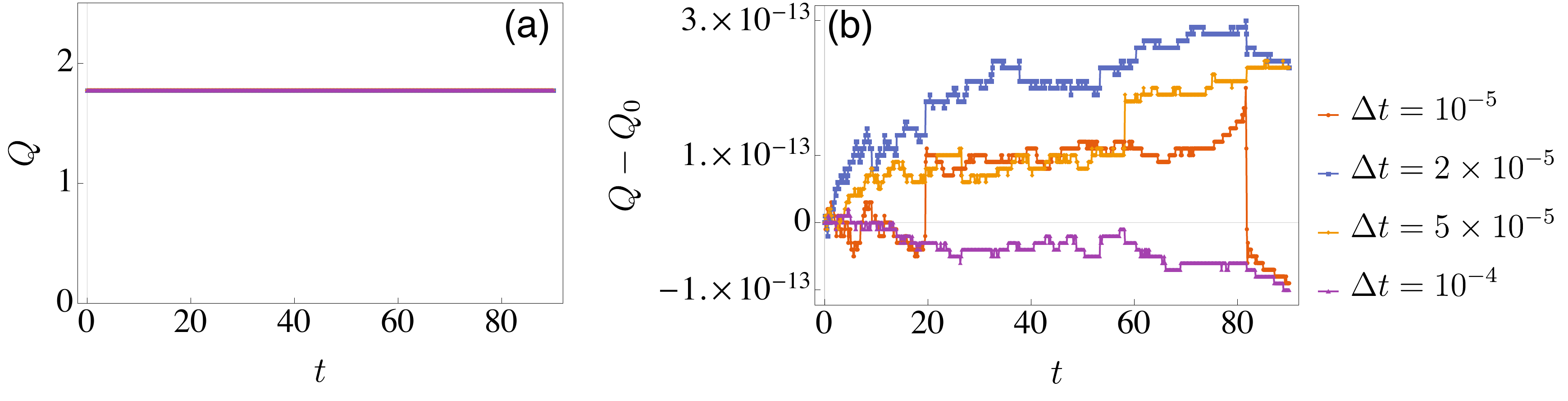}
\caption{The conserved total number of right and left movers, $Q=N_R+N_L$ (a) 
and 
$Q - Q(t=0)$ (b) in the dynamics.
Here $n_0=1$, $\xi=1$, $W=1.0$ and $T_0=0.01$.
}
\label{fig:Qt}
\end{figure}

\begin{figure}[b]
\centering
\includegraphics[width=0.8\textwidth]{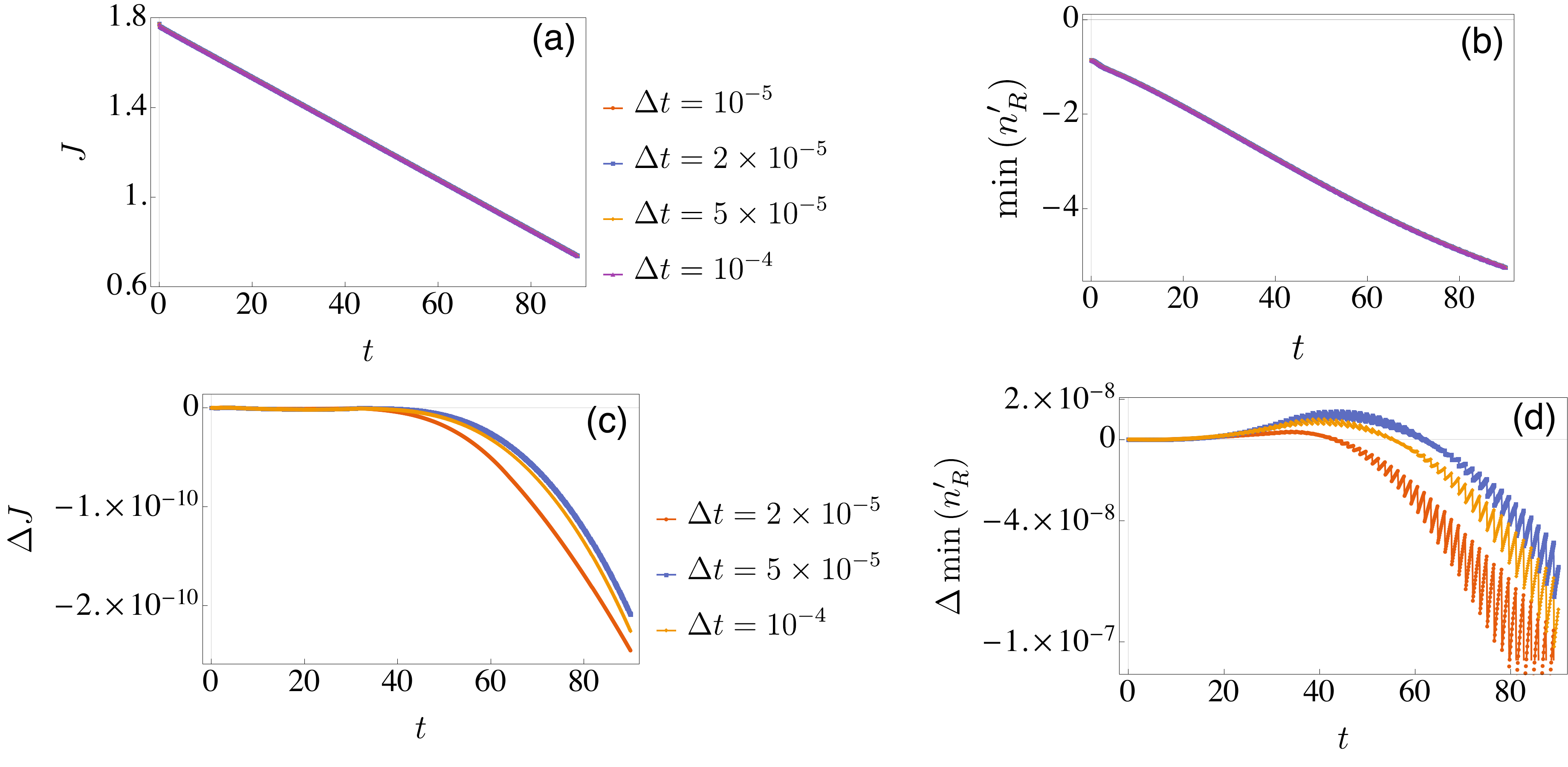}
\caption{The total current $J = N_R - N_L$ (a), as well as 
the minimal value of the slope (along the shock front) for the right movers, 
$\min{n_R'}$ (b), both plotted versus time $t$. Both are approximately linear
functions of time. 
The bottom panels (c) and (d) show the variation of these quantities
$\Delta J$ and $\Delta \min{n_R'}$
induced by changes in the time step $\Delta t$ for the Runge-Kutta. 
These show that our results are well-converged with respect to the time step. 
Here $n_0=1$, $\xi=1$, $W=1.0$ and $T_0=0.01$.
}
\label{fig:dnRI}
\end{figure}

The total charge 
$Q \equiv N_R + N_L$ 
is conserved in the dynamics, which is one check for the accuracy of our numerics. 
In Fig.~\ref{fig:Qt}, $Q(t)$ is shown for several different choices of the time step, indicating conservation
to high accuracy. 
In the dynamics, the total current 
$J \equiv N_R - N_L = 2 N_R - Q$ of right- and left-moving fermions decreases linearly in time
[Fig.~\ref{fig:dnRI}(a)], which allows us to define the decay time $t_\textsf{decay}$
\begin{equation}
	\frac{\Delta N_R(t)}{N_R(0)}
	=
	\alpha
	\frac{t}{t_\textsf{decay}},
	\label{eq:definetdecay}
\end{equation}
where $\alpha < 1$ is an arbitrary numerical factor used to set the threshold decay level
for $\Delta N / N$. 

On the other hand, a shock front develops in the dynamics, 
which can be characterized by the negative minimum value of the slope in $n_R(x)$, 
defined as $\min(n'_R)$. From Fig.~\ref{fig:dnRI}(b), we see that $\min{(n'_R)}$ also decreases (approximately) 
linearly in time, leading to the definition of the slope time $t_\mathsf{slope}$,
\begin{equation}
	\min{[n'_R(t)]} 
	\simeq 
	-
	\frac{n_0}{\xi}
	\left(
		1
		+
		\frac{t}{t_\mathsf{slope}}
	\right).
\label{eq:minnR'}
\end{equation}

In Figs.~\ref{fig:dnRI}(c) and (d), we also show the difference of $J(t)$ and $\min{n'_R(t)}$ 
at $\Delta t=10^{-4}$ with that of several different time steps $\Delta t$. 
It can be seen that the quantities are converged with respect to the time step. 

\begin{figure}
\centering
\includegraphics[width=0.6\textwidth]{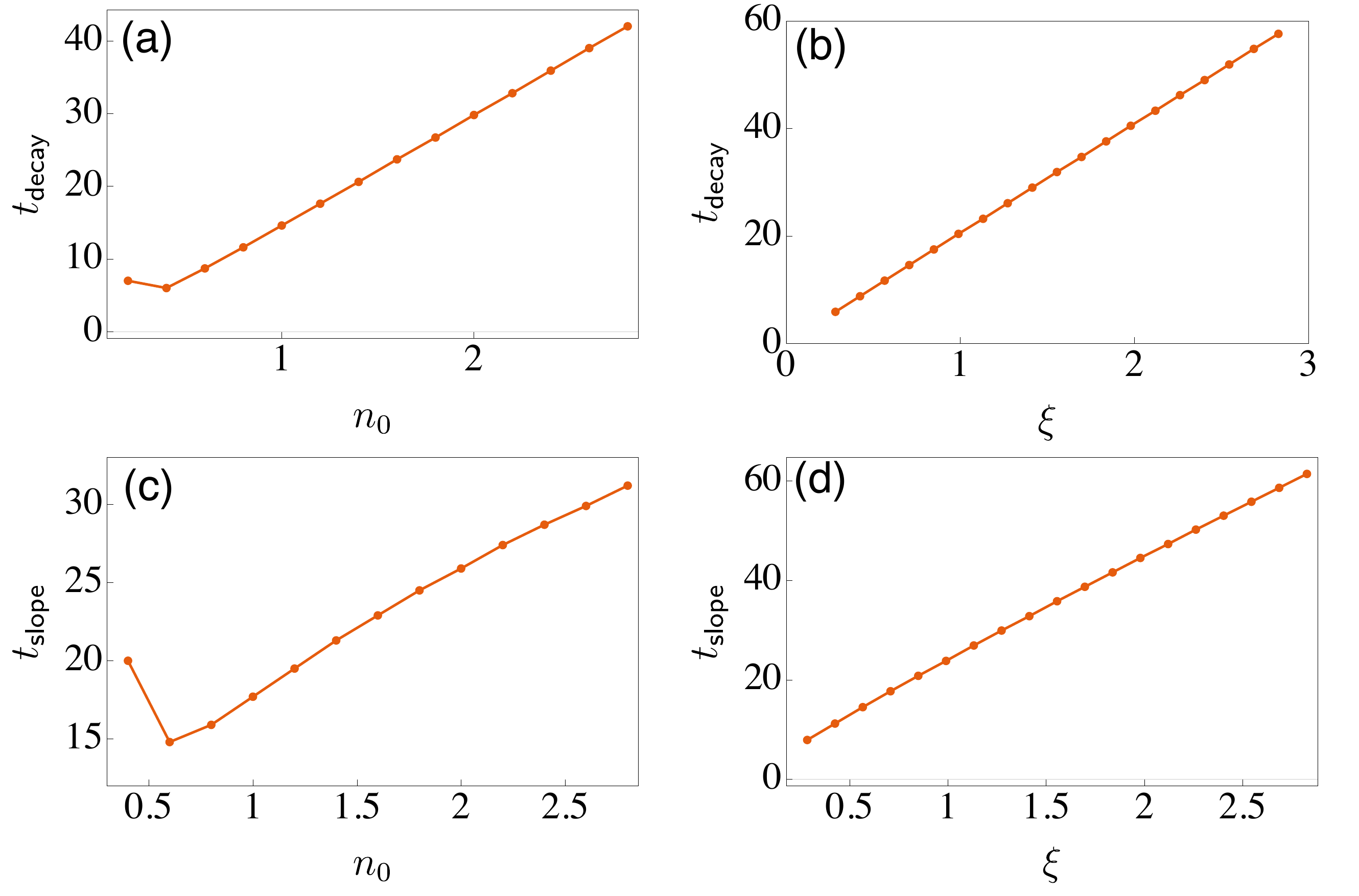}
\caption{The decay time $t_\textsf{decay}$ and shock wave development time $t_\mathsf{slope}$ as a function of the initial peak height 
$n_0$ and width $\xi$ [Eq.~(\ref{eq:nR0x})]. 
Here $W=1.0$ and $T_0=0.01$, $\xi=1$ for varying $n_0$ and $n_0=1$ for varying $\xi$.}
\label{fig:tn0xi}
\end{figure}

The dependence of $t_\textsf{decay}$ and $t_\mathsf{slope}$ on the parameters $n_0$ and $\xi$
is exhibited in Fig.~\ref{fig:tn0xi}. The bound given by Eq.~(14) in the main text for 
$t_\textsf{decay}$ is 
\begin{align}
	t_\textsf{decay}
	=
	\alpha
	\frac{\sqrt{3 \pi} n_0 \xi}{T_0}.
\end{align}
This bound corresponds to cooling the emitted left movers to absolute zero, $T_L = 0$. 
The linear dependence on $n_0$ and $\xi$ is confirmed in Fig.~\ref{fig:tn0xi}(a,b), 
while the Fig.~3(a) in the main text shows the variation with the initial temperature $T_0$. 
The shock-slope time $t_\mathsf{slope}$ behaves similarly, except the latter continues to exhibit
$W$-dependence in the strong-coupling $W \gtrsim 1$ limit [Fig.~3(d) in the main text]. 
Given that the emitted left movers, whose density $n_L \lesssim T_0/\sqrt{12}$ 
maintains the quasi-equilibration in the bulk of the packet, originate
from the hot leading edge of the shock front, one expects that 
$t_\mathsf{slope}$ is of the same order as $t_\textsf{decay}$, with similar
dependence on $n_0$, $\xi$, and $T_0$.

\subsection{Short-time Perturbation Theory}\label{sec:perturbation}

In this section, we consider the initial relaxation dynamics induced
by the imbalance force $I$ [Eq.~(12) in the main text] for the 
right-moving density excess. The analysis holds only in a transient window
of duration 
\begin{align}\label{tiDef}
	t \lesssim t_{\mathsf{i}} \equiv v_F / (W^2 n_0^3),
\end{align} 
where $n_0$ is the amplitude of the excess [Eq.~(\ref{eq:nR0x})] induced
by the laser pulse.  
At $t = 0$, the left-mover density $n_L = 0$ everywhere. As a result,
$I < 0$ throughout the right-mover excess. This immediately induces
the conversion of right to left movers, wherein $n_L(t,x)$ acquires
a positive density over the range $\left(-v_F t,v_F t\right)$
and $T_{L}$ decreases in the same region (in order to keep $P_{L}$ unchanged).
On the other hand, the excess right-mover density decreases, while 
its temperature increases [Eq.~(\ref{eq:TRLtx})]. The density and temperature of 
the right movers approach zero exponentially away from the center of the wave packet.
We always have $\{n_{L},T_{L}\} \ll \{n_R, T_R\}$, which is confirmed by the numerics.
Therefore we can ignore $n_{L}$ and $T_{L}$ terms in the last factor of Eq.~(12)
in the main text, leading to ($v_F = k_B = 1$) 
\begin{equation}
	I
	\simeq 
	-
	\frac{\pi^{2}W^{2}}{96} 
	\left( 
		\frac{1}{e^{-2\pi n_R/T_R}+1} 
		- 
		\frac{1}{e^{-2\pi n_L/T_L}+1} 
	\right) 
	\left(4 n_{R}^{2} + T_{R}^{2}\right)
	\left(4 n_{R}^{2} + 9 T_{R}^{2}\right).
\end{equation}

\begin{figure}[b!]
\centering
\includegraphics[width=0.4\textwidth]{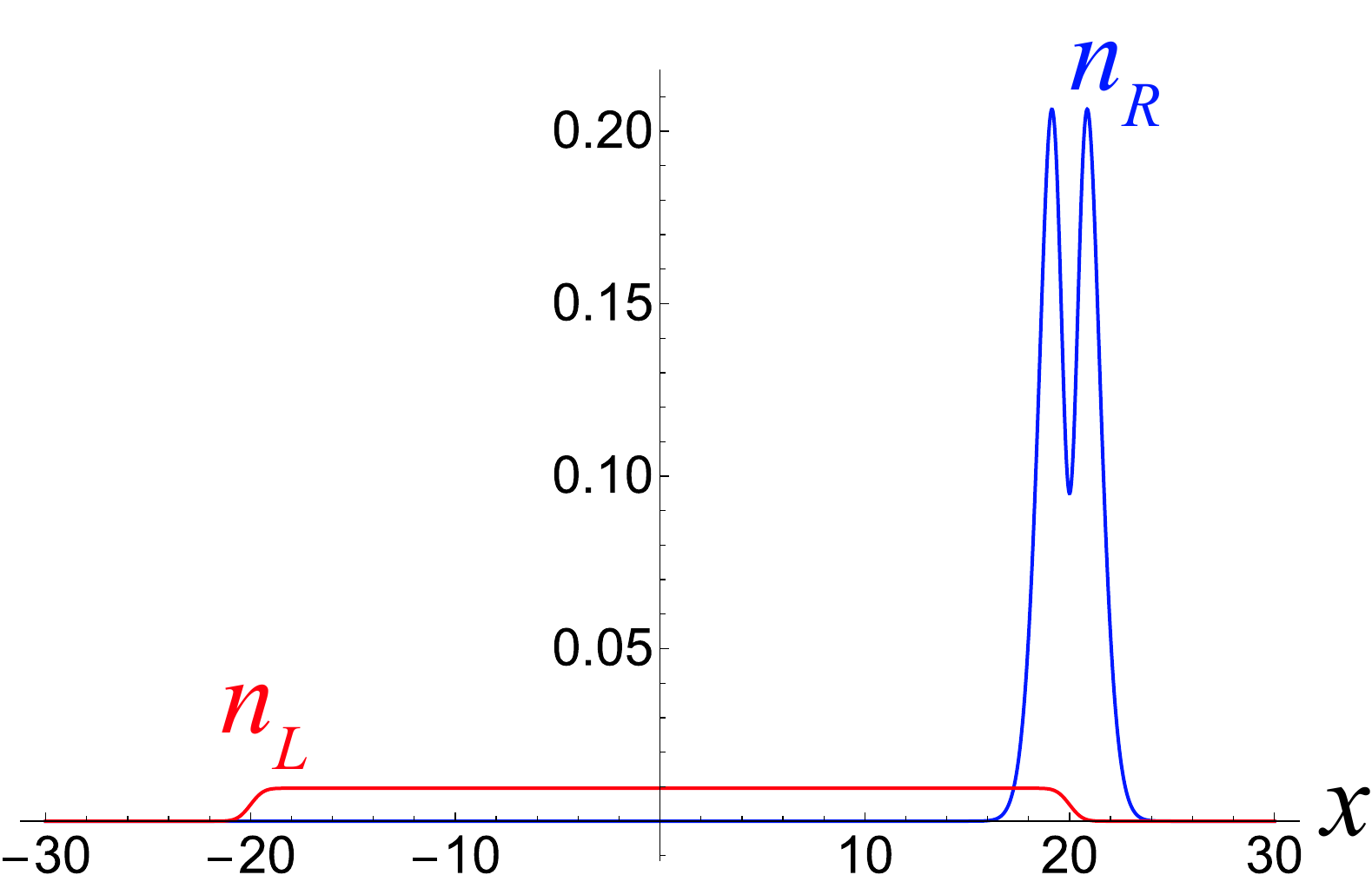}
\caption{Transient (short-time) dynamics predicted by perturbation theory for $n_{R,L}(t,x)$, 
Eq.~(\ref{PTResults}). Lowest-order perturbation theory fails to capture the shock formation,
because it does not incorporate the feedback of the density conversion into the 
temperature profiles [Eq.~(\ref{eq:TRLtx})]. In the numerical solution to the full 
equations of motion shown in Fig.~2 of the main text, this feedback effect quickly suppresses
the imbalance $I$ throughout the bulk of the right-moving excess, leading to a ``quasi-equilibration''
there. The shock forms at the leading hot edge precisely because the balance in the bulk can only 
be sustained by continually emitting left movers with quasi-stationary amplitude $n_L \lesssim T_0/\sqrt{12}$
[the equal sign corresponding to cooling of the left movers to absolute zero].  
}
\label{fig:PT}
\end{figure}

Then we have 
\begin{align}
	n_{R,L}\left(t, x\right) 
	=&\,
	n_{R,L}^{\text{\tiny (0)}}\left(x \mp t\right)
	\pm
	\int_{0}^{t}dt'I\left[t', x \mp \left(t-t'\right)\right].
\end{align}
At first order, we can approximate the density and temperature in the integral by the solution from the homogeneous equations,
\begin{equation}
	n_{R,L}\left(t, x\right)
	=
	n_{R,L}^{\text{\tiny (0)}}\left(x \mp t\right),
\qquad 
	P_{R,L}\left(t,x\right)
	=
	P_{R,L}^{\text{\tiny (0)}}\left(x \mp t\right).
\end{equation}
The expression for $P_{R,L}$ is exact (holds to all orders) when there is no electric force.
For the initial conditions in Eq.~(\ref{eq:nR0x}) with $T_R(0,x) = T_L(0,x) = T_0$, 
we can make approximation that 
$T_{R,L}\left(t,x\right) \simeq T_{0} \ll n_R$ at first order. 
Then we have 
\begin{subequations}
\begin{align}
	I\left[x - \left(t-t'\right),t'\right] 
	\simeq&\,
	-
	\frac{\pi^{2}W^{2}}{12}
	\left[n_{R}^{\text{\tiny (0)}}\left(x - t\right)\right]^{4},
\\
	I\left[x + \left(t-t'\right),t'\right] 
	\simeq&\,
	\frac{\pi^{2}W^{2}}{12}
	\left\{n_{R}^{\text{\tiny (0)}}\left[x + \left(t - 2t'\right)\right]\right\}^{4},
\end{align}
\end{subequations}
leading to [via Eq.~(\ref{eq:nR0x})]
\begin{subequations}\label{PTResults}
\begin{align}
	n_{R}\left(t,x\right) 
	\simeq&\,
	n_{R}^{\text{\tiny (0)}}\left(x - t\right)
	-
	\frac{\pi^{2}W^{2}}{12}
	\,
	t
	\left[n_{R}^{\text{\tiny (0)}}\left(x-t\right)\right]^{4}
\nonumber\\
	\simeq&\,
	n_0 \, e^{-(x - t)^2/\xi^2} 
	- 
	\frac{\pi^2 W^2 n_0^4}{12} 
	\,
	t 
	\,
	e^{-4(x -t)^2/\xi^2},
\\
	n_{L}\left(t,x\right) 
	\simeq&\,
	\frac{\pi^{2}W^{2}}{12}
	\int_{0}^{t} dt' \left[n_{R}^{\text{\tiny (0)}}\left(x + \left(t-2t'\right)\right)\right]^{4}
\nonumber \\
	\simeq&\,
	\frac{\pi^{5/2}W^{2}\xi n_0^4}{96}
	\left\{
		\erf\left[2\left(\frac{t - x}{\xi}\right)\right]
		+
		\erf\left[2\left(\frac{t + x}{\xi}\right)\right]
	\right\}.
\end{align}
\end{subequations}
Here $\erf(x)$ is the error function.
The density change is 
\begin{align}
	\frac{\Delta N_R}{N}
	=
	-
	\frac{\pi^{3/2}}{24} 
	\left(W^2 n_0^3\right)
	t,
\end{align}
which determines the transient time $t_{\mathsf{i}}$ [Eq.~(\ref{tiDef})]. 
The perturbative results are depicted in Fig.~\ref{fig:PT}.

\subsection{Equations of Motion for the Distribution Functions}

The coupled equations of motion for the densities $n_{R,L}(t,x)$  [Eq.~(11) in the main text] 
are complicated due to the strong (exponential) nonlinearity of the imbalance force
in these variables, Eq.~(12) in the main text. 
At the same time, the relatively simple behavior of the distribution functions $f_{R,L}^0$ throughout
the ``quasi-equilibration'' region of the right-moving density excess 
[where the imbalance is zero, Figs.~2(e,g) in the main text], as well as in the shock region, 
suggests that it could prove easier to analyze the dynamics of these functions. 
This is what we attempt to do in this section.

We continue to set $v_F = k_B = 1$. 
We define 
\begin{align}
	g_{R,L}
	\equiv
	1 
	-
	f_{R,L}^0
	=
	\left[
	e^{2 \pi n_{R,L} / T_{R,L}} 
	+
	1
	\right]^{-1}.
\end{align}
We introduce the relative temperatures 
\begin{align}
	\ton_{R,L}
	\equiv
	\frac{T_{R,L}}{n_{R,L}}
\end{align}
and the normalized initial density profile 
\begin{align}
	\nrz 
	\equiv 
	n_{R}^{\text{\tiny (0)}}/T_0
\end{align}
[see Eqs.~(8) and (10) in the main text].
In terms of the distribution functions $g_{R,L}$, 
the equations of motion (11) and (12) can be 
exactly rewritten as follows, 
\begin{align}
\label{fRfLEOM}
\!\!\!\!
\begin{aligned}
\phantom{0}
\\
	\sqrt{
	1
	+
	12
	\left[\nrz(x_-)\right]^2
	}
	\left[
	\frac{
	\ton_R^3
	}{
	\left(
	12
	+
	\ton_R^2
	\right)^{3/2}
	}
	\right]
	\!
	\parr_+
	g_R
	=&\,
	-
	\left(
	\pi \lambda
	\right)
	\,
	g_{R}
	\left[
		1 - g_{R}
	\right]
	\left[
	g_R
		-
	g_L
	\right]
	\left[
	\left\{
	1
	+
	12
	\left[
		\nrz(x_-)
	\right]^2
	\right\}^2
	\Xi(\ton_R)	
	+
	\Xi(\ton_L)	
	\right]\!,
\\
	\left[
		\frac{\ton_L^3}{\left(12 + \ton_L^2\right)^{3/2}}	
	\right]
	\!
	\parr_-
	g_L
	=&\,
	-
	\left(
	\pi \lambda
	\right)
	\,
	g_{L}
	\left[
		1 - g_{L}
	\right]
	\left[
	g_R
		-
	g_L
	\right]
	\left[
	\left\{
	1
	+
	12
	\left[
	\nrz(x_-)
	\right]^2
	\right\}^2
	\Xi(\ton_R)	
	+
	\Xi(\ton_L)	
	\right]\!.
\\
\phantom{0}
\end{aligned}
\!\!\!\!\!\!\!
\end{align}
In these equations, 
\begin{align}
	\lambda
	\equiv
	\pi^2 W^2 T_0^3/96,
\end{align}
and the function $\Xi(x)$ simply interpolates between the
finite bounds 1/9 and 9, 
\begin{align}\label{XiDef}
	\Xi(x)
	\equiv
	\left(4 + x^2 \right)
	\left(4 + 9 \, x^2\right)
	\left(12 + x^2 \right)^{-2}.
\end{align}

\begin{figure}[th!]
\begin{center}
\includegraphics[width=0.71\textwidth]{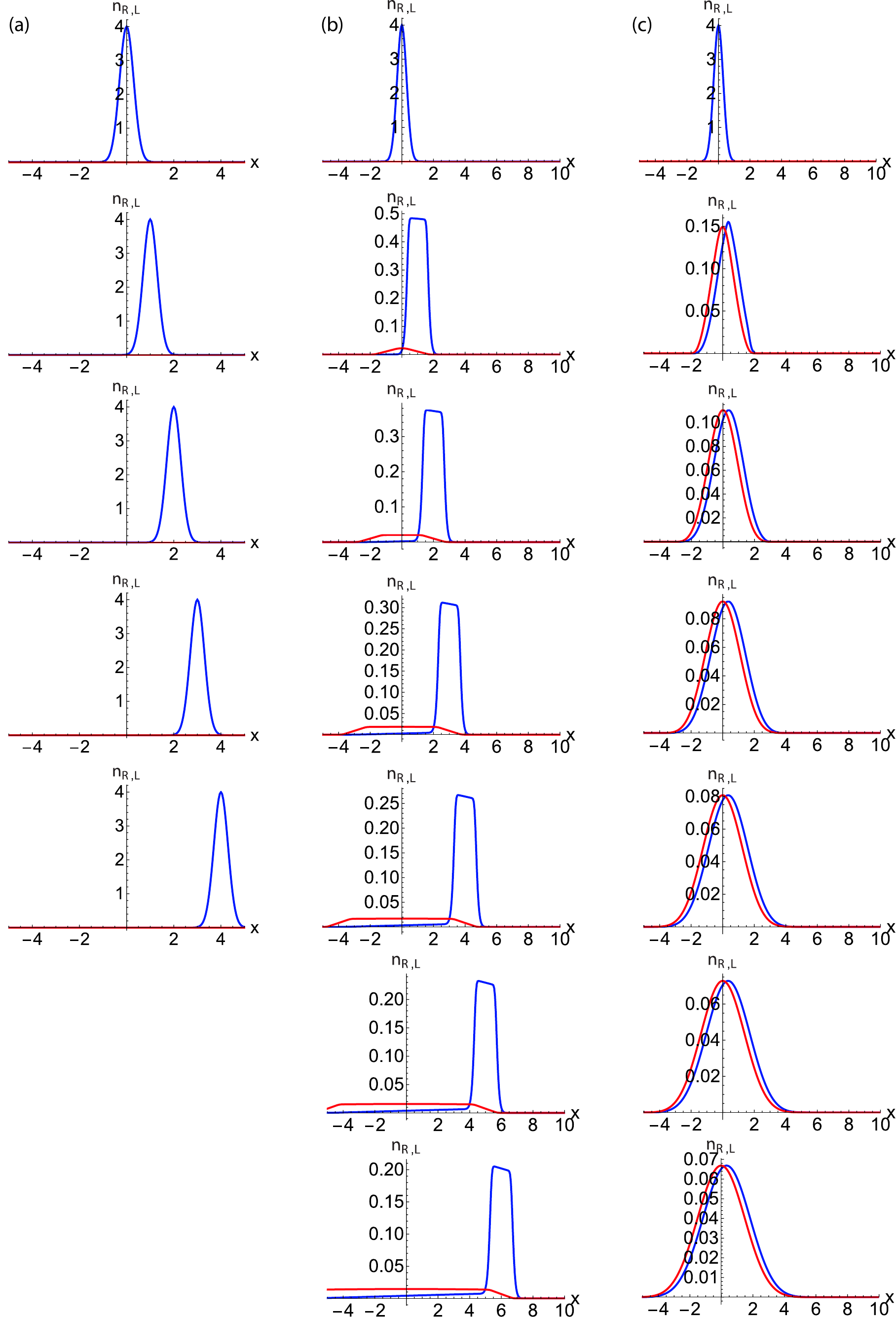}
\caption{ 
Numerical integration of the Green's function solution Eq.~(\ref{GFSol}) to the simplified EOM (\ref{EOM-Simp}) for the distribution functions $g_{R,L}$.
Results are re-expressed in terms of $n_{R,L}(t,x)$. 
Subpanel (a): vanishing mass $m = 0$. 
(b): small mass $m = 0.1$. 
(c): large mass $m = 3$. 
The \blue{blue} trace indicates $n_R(t,x)$, while the \darkred{red} trace is $n_L(t,x)$. 
All cases have the same initial condition $\nrz(x) = n_0 \exp(-x^2/2\xi^2)$, with $n_0 = 4$ and $\xi = 0.3$. 
Panel (c) indicates that the total number density is \emph{not} conserved by the simplified distribution function Eqs.~(\ref{EOM-Simp}).
The phenomenology of (b) is somewhat similar to the real dynamics (Fig.~2 in the main text), except that there is a strong distortion and flattening 
of the initial right-mover peak. (c) shows an effective equilibration between $n_{R,L}$, which both appear to slowly diffuse around the original 
center-of-mass position $x = 0$. This behavior is very different from the relativistic propagation of the right-moving packet observed in numerics,
but does demonstrate the effect of feedback that is neglected in the perturbation theory, Fig.~\ref{fig:PT}.  
\label{Fig--MassPlots}}
\end{center}
\end{figure}

\subsubsection{Toy Version}

Consider the simplified, linear equations
\begin{align}\label{EOM-Simp}
\begin{aligned}
	\parr_+
	g_R
	=&\,
	-
	\lambda
	\left(
	g_R
		-
	g_L
	\right),
\\
	\parr_-
	g_L
	=&\,
	-
	\lambda
	\left(
	g_R
		-
	g_L
	\right).
\end{aligned}
\end{align}
These can be cast in the form of a matrix differential equation,
\begin{align}
	\begin{bmatrix}
	\parr_+ + \lambda 	&	- \lambda		 \\
	-\lambda			& 	- \parr_- + \lambda
	\end{bmatrix}
	\begin{bmatrix}
		g_R 
		\\
		g_L
	\end{bmatrix}
	=
	\begin{bmatrix}
	(\parr_t +   \parr_x) 		+ 2   \lambda 			&	- 2   \lambda		 \\
	-2   \lambda								& 	(\parr_t -   \parr_x) + 2   \lambda
	\end{bmatrix}
	\begin{bmatrix}
		g_R 
		\\
		g_L
	\end{bmatrix}
	=
	0,
\end{align}
or 
\begin{align}\label{DiracEq}
	\left[
	\bigg(
		i \parr_t 
		+
		\sigh^3
		i \parr_x
	\bigg)
		-
		i m
		\,
		\sigh^1
		+
		i m 
	\right]
	f
	=
	0,
\qquad
	m
	=
	2   \lambda.
\end{align}
This is a massive 1+1-D Dirac equation, with imaginary (tachyonic) mass $M = i m$ and scalar potential $V = -i m$. 
To compute the evolution from an initial condition $g_0(x)$, we construct the Green's function for Eq.~(\ref{DiracEq}):
\begin{align}
	\left[
	\bigg(
		i \parr_t 
		+ 
		\sigh^3
		i \parr_x
	\bigg)
		-
		i m
		\,
		\sigh^1
		+
		i m 
	\right]
	G(t,x)
	=
	\delta(t) \, \delta(x).
\end{align}
The solution is \cite{SM--Tim2011}
\begin{align}
\label{GResult}
	G(t,x)
	=
	-
	i
	\theta(t)
	\,
	e^{- m t}
\left\{
		\begin{bmatrix}
		\delta(t-x)	&	0\\
		0		&	\delta(t+x)
		\end{bmatrix}
		+
		\begin{bmatrix}
		G^\pup{1}(t,x;m) 	& G^\pup{2}(t,x;m)\\
		G^\pup{2}(t,x;m) 	& G^\pup{1}(t,-x;m)
		\end{bmatrix}
		\theta\left(t^2 - x^2\right)
\right\},
\end{align}
where
\begin{align}
	G^\pup{1}(t,x;m)
	=&\,
	\frac{m}{2} 
	\left(\frac{t + x}{\sqrt{t^2 - x^2}}\right)
	I_1\left(m \sqrt{t^2 - x^2}\right),
\quad
	G^\pup{2}(t,x;m)
	=
	\frac{m}{2} 
	I_0\left(m \sqrt{t^2 - x^2}\right).
	\label{BesselIComp}
\end{align}
In these equations, $I_\nu(z) = i^{-\nu} \, J_{\nu}(i z)$ denotes the modified Bessel function of the first kind.

In the large-argument limit, 
\[
	\lim_{x \rightarrow \infty}
	I_\nu(x) 	
	\rightarrow
	\frac{1}{\sqrt{2 \pi x}}
	e^x
	\left[1 + \ord{\frac{1}{x}}\right].
\]
Such exponential growth is countered in Eq.~(\ref{GResult}) 
by the prefactor $e^{- m t}$ (which arose from the imaginary scalar potential). 
The initial condition for the quench is encoded in 
\begin{align}
	g_0(x)
	=
	\begin{bmatrix}
	g_R^\pup{0}(x)
	\\
	g_L^\pup{0}(x)
	\end{bmatrix}
	\equiv
	\begin{bmatrix}
	\left\{\exp\left[2 \pi \nrz(x)\right] + 1\right\}^{-1}
	\\
	1/2
	\end{bmatrix},
\end{align}
such that $g_R^\pup{0} \ll 1$ throughout the bulk of the initial packet, but equal to $1/2$ everywhere else. 

The time evolution is given by 
\begin{align}\label{GFSol}
	g(t,x)
	=
	\begin{bmatrix}
		g_R(t,x) 
	\\	
		g_L(t,x)
	\end{bmatrix}
	=&\,
	\int_{-\infty}^{\infty}
	d x'
	\,
	i
	\,
	G(t,x')
	\,
	g_0(x - x')
\nonumber\\
	=&\,
	e^{- m t}
\left\{
		\begin{bmatrix}
			g_R^\pup{0}(x -	t)		
		\\	
			1/2 
		\end{bmatrix}	
		+
	\int_{-t}^{t}
	d x'
		\begin{bmatrix}
			G^\pup{1}(t,x';m) \, g_R^\pup{0}(x - x') +  \frac{1}{2}G^\pup{2}(t,x';m)
		\\
			G^\pup{2}(t,x';m) \, g_R^\pup{0}(x - x') +  \frac{1}{2}G^\pup{1}(t,x';m)
		\end{bmatrix}
\right\}.
\end{align}
Although it is not immediately obvious, this equation preserves $g_R = g_R^\pup{0} = 1/2$,
the case of homogeneous equilibrium. 
The results for vanishing, small, and large ``mass'' $m$ are depicted in Fig.~\ref{Fig--MassPlots}.

\section{Dissipative Interpacket Collisions: Bounces and Current Switching}

\begin{figure}[th!]
\begin{center}
\includegraphics[width=0.95\textwidth]{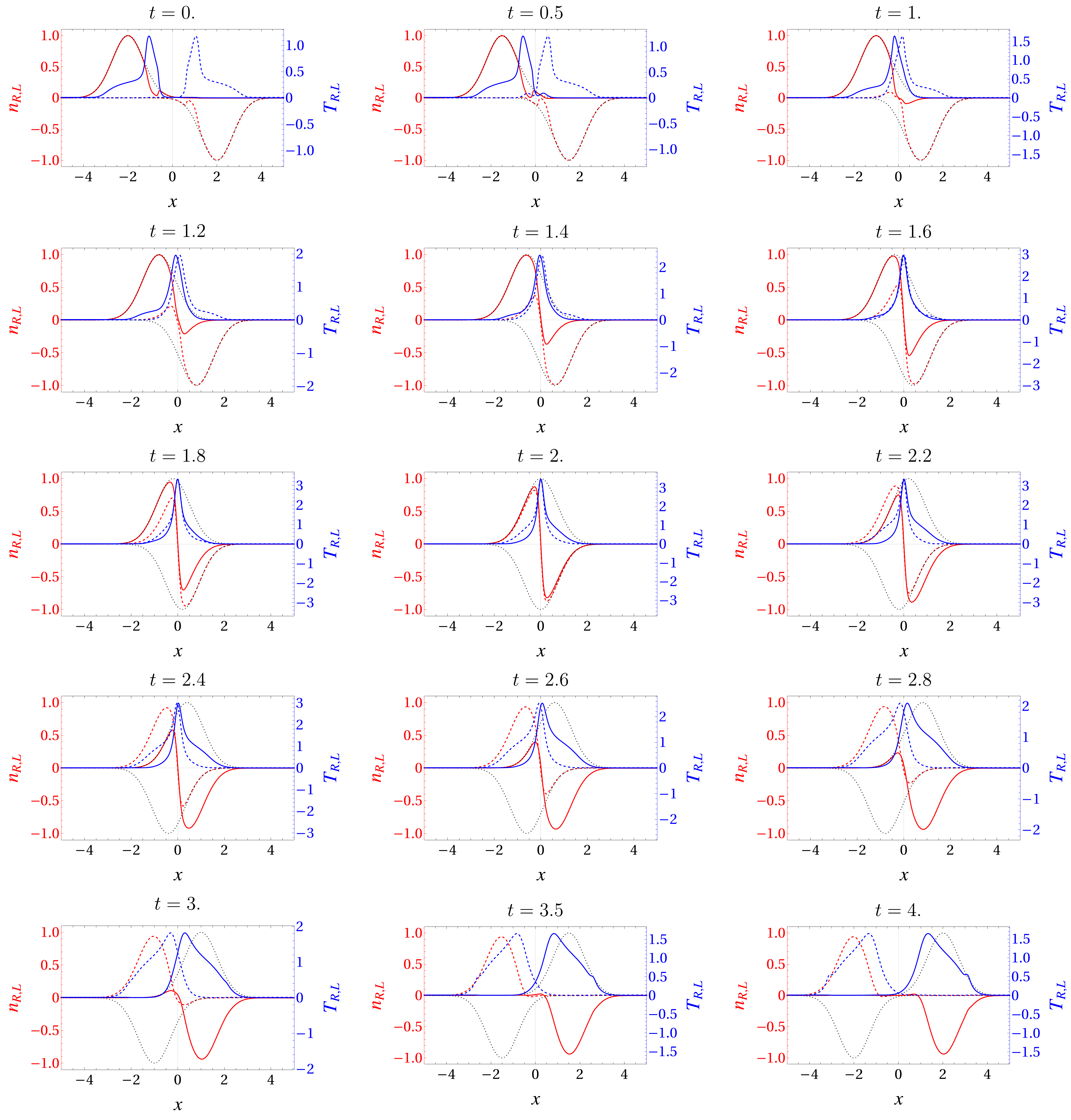}
\caption{The densities and temperatures of right- and left-moving packets 
for an oppositely-charged packet collision, with $W=1.0$ and $T_0=0.01$. 
The right-mover (left-mover) density $n_R$ ($n_L$) is depicted by the solid (dashed)
\darkred{red} line. 
The right-mover (left-mover) temperature $T_R$ ($T_L$) is depicted by the solid (dashed)
\blue{blue} line. 
The black dotted curves depict the ballistic propagation of packets without scattering (rigid shifting).
Initially, the hot spots of the right and left movers reside at the shock fronts. 
As the two packets move close to each other, the collision center heats up further and the density of each packet 
flips sign around the collision center. The packets change the sign of the density after passing through each other, 
but otherwise approximately maintain their shapes. The bouncing dynamics for the densities are very similar to that of two solitons,
except that here the collision results in significant heating. The heating in subsequent collisions is minimal, and does not
lead to a rapid dissipation of the current, see Fig.~4(d) in the main text, and Figs.~\ref{fig:Current-W} and \ref{fig:Current-W-n0=0.1}, below.
Despite the heating throughout the packets due to the collision, the imbalance remains small everywhere except near the collision center,
and along the shock fronts as the packets approach or propagate away from the center, see Fig.~\ref{fig:nI_W=1}.} 
\label{fig:nT_W=1}
\end{center}
\end{figure}

\begin{figure}[th!]
\begin{center}
\includegraphics[width=0.95\textwidth]{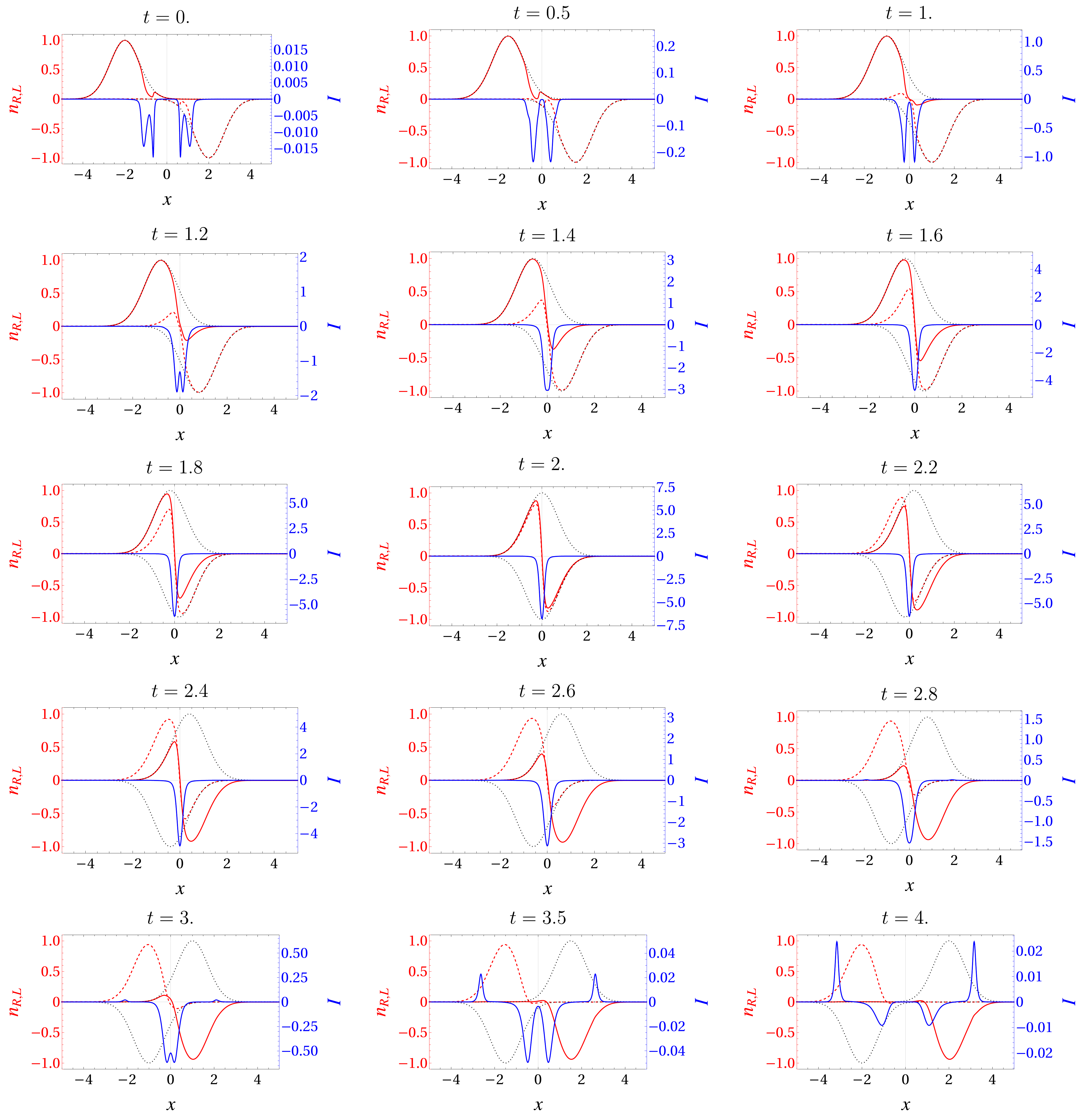}
\caption{The same as Fig.~\ref{fig:nT_W=1}, but showing the densities (\darkred{red}) and the imbalance $I$ (\blue{blue}) through the collision. 
At $t=0$, the imbalance peaks around the two shock fronts of the right- and left-moving packets. 
During the collision, the imbalance peaks around the collision center, where the temperature is highest and the imbalance at $t=2$ is several 
orders-of-magnitude larger than that at the shock fronts. 
This central hot spot mediates the collision, but the imbalance remains negligible away from it (signaling quasiequilibration throughout
the bulks of the packets, as similarly found in the shock formation mechanism). 
After passing through each other, the imbalance again begins to peak around the new shock fronts at $t=4$.} 
\label{fig:nI_W=1}
\end{center}
\end{figure}

\begin{figure}[th!]
\begin{center}
\includegraphics[width=0.95\textwidth]{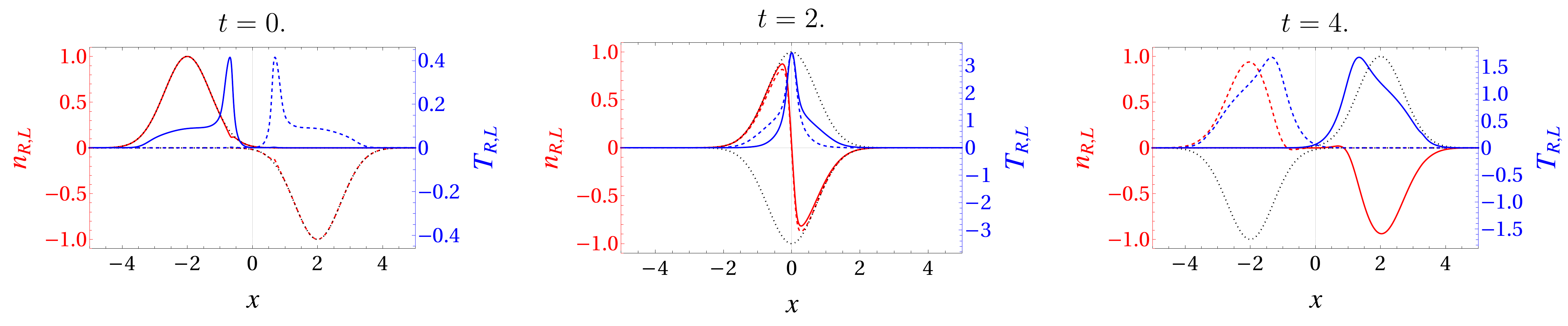}
\caption{The density and temperature profiles of the right and left moving packets 
with $T_0=0.001$ and $W=1$, \textit{cf.} Fig.~4 in the main text.}
\label{fig:nT_T0=0.001}
\end{center}
\end{figure}

\begin{figure}[th!]
\begin{center}
\includegraphics[width=0.95\textwidth]{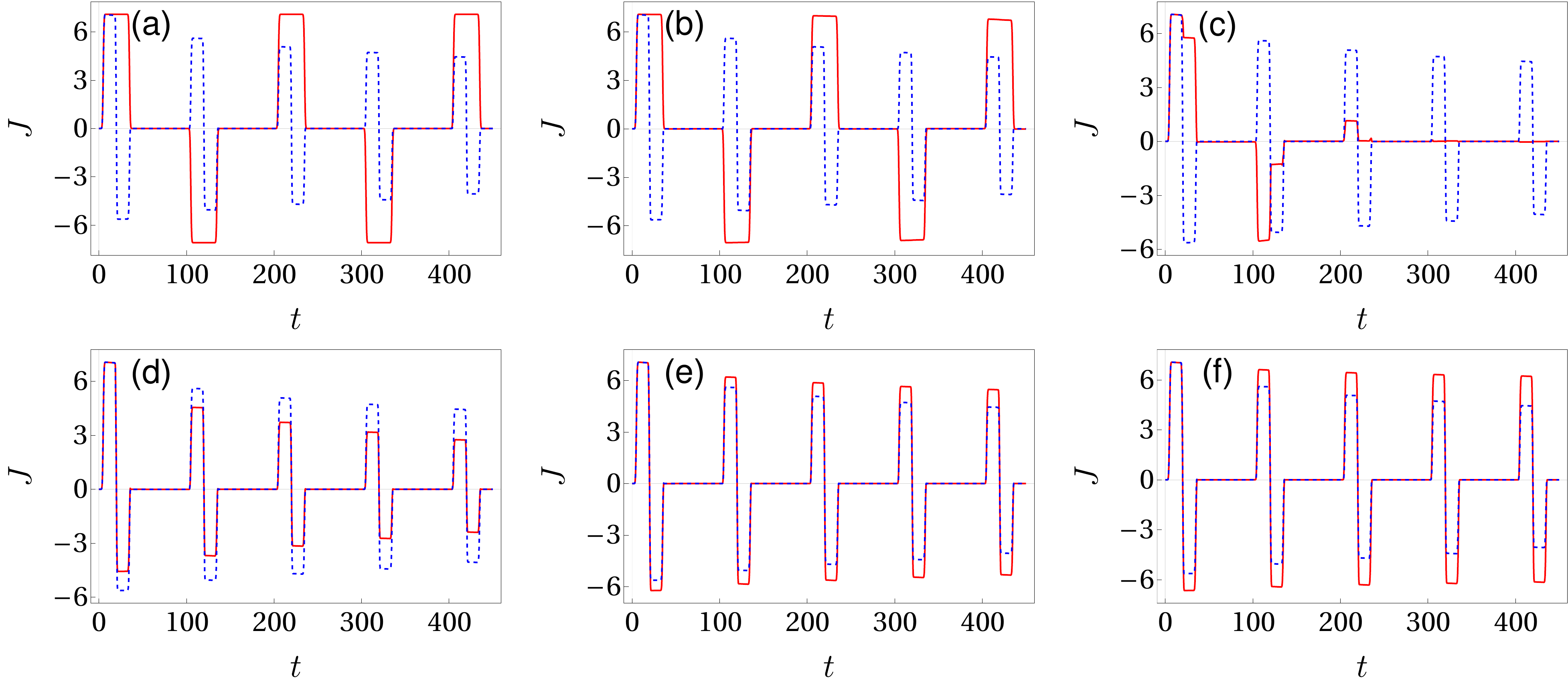}
\caption{The current along the $y$-direction of a rectangular sample with $n_0=1$, $\xi=1$, initial temperature $T_0=0.001$, 
and different 1PU interaction strengths $W$. We take the $W = 1$ case (the blue dashed curve) as a reference. 
(a): $W=0$. (b): $W=0.01$. (c): $W=0.1$. (d): $W=0.5$. (e): $W=2$. (f): $W=5$.}
\label{fig:Current-W}
\end{center}
\end{figure}

\begin{figure}[th!]
\begin{center}
\includegraphics[width=0.95\textwidth]{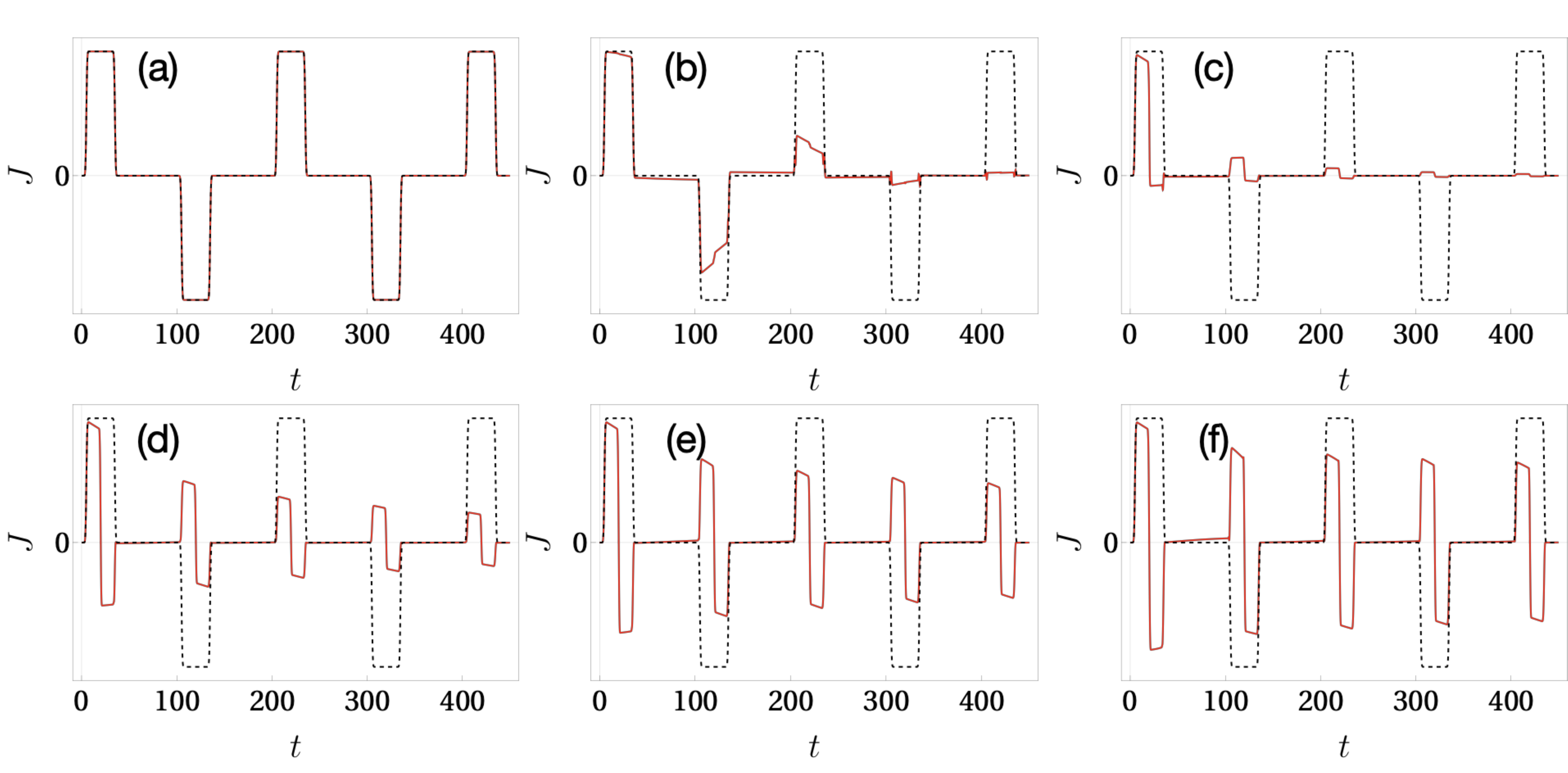}
\caption{The current along the $y$-direction of a rectangular sample with $n_0=0.1$, $\xi=1$,  initial temperature $T_0=0.001$, 
and different 1PU interaction strengths $W$. 
The $W=0$ case (the black dashed curve) is plotted as a reference. 
(a): $W=0.1$. (b): $W=1$. (c): $W=5$. (d): $W=10$. (e): $W=20$. (f): $W=50$.}
\label{fig:Current-W-n0=0.1}
\end{center}
\end{figure}

Now we consider the dynamics of two packets of right and left movers. 
When the two packets have the same sign of the density, they pass through each other nearly ballistically at low temperature,
since the 1PU scattering is strongly suppressed. 
The nontrivial dynamics occurs when we consider collisions between packets with opposite densities, e.g.,
one right-moving packet with a positive-density excess 
and 
one left-moving one with a negative-density deficit,
\begin{equation}
	n_R(x,t=0)
	= 
	n_0 
	\,
	e^{-(x - x_1)^2/\xi^2},
	\qquad 
	n_L(x,t=0) 
	= 
	-
	n_0 
	\,
	e^{-(x - x_2)^2/\xi^2},
\label{eq:nRL0}
\end{equation}
with $x_1 < 0$ and $x_2 > 0$. 
Here we consider packets with identical shapes and opposite profiles, since this is what is generated in the ideal
case by the electric-pulse quench described in Fig.~1 of the main text and considered in Sec.~\ref{sec:SM--Quench}, above. 
The near-perfect retroreflection described below is not significantly modified for small violations of this parity (changes
in total density of right vs.\ left movers and/or deformations of their shape profiles). 

In Figs.~\ref{fig:nT_W=1} and \ref{fig:nI_W=1}, we show the 
densities and temperatures for the right and left movers, as well as the imbalance $I$,
for a collision of oppositely-charged packets with 
$T_0=0.01$ and $W=1$. 
Initially, the dissipative evolution of the separate packets is driven by the shock-forming mechanism described in the main 
text and in Fig.~2. 
The imbalance $I$ peaks at the hot spots riding the shock fronts of the impinging packets.  
As the two packets collide with each other, the collision center heats up further and becomes the new out-of-equilibrium hot spot. 
The strongest imbalance peak for the collision is several orders-of-magnitude 
larger than that for the shock fronts. The right movers convert to left movers rapidly and the packets flip their signs after passing through each other. 
Then the packets separate, and subsequently develop new shock fronts. 
In Fig.~\ref{fig:nT_T0=0.001}, we show the density and temperature of the right- and left-moving packets with 
$T_0=0.001$, as in Figure 4 of main text. While the shock evolution is slower at lower temperatures, the near-perfect retroreflection 
improves with decreasing $T_0/n_0$. This can be understood from the unique form of the 1PU scattering imbalance force [Eq.~(12) of the main text], 
which is not suppressed if right and left movers have opposite densities, even as $T_0 \rightarrow 0$. 

On a rectangular sample with packets induced on both right and left $y$-edges by the initial electric-field quench, the current on the $y$-axis is 
nonzero only when the packets propagate along the $y$-edges, while the $x$-axis current always vanishes due to the cancellation of the right- and left-moving packets
[see Figs.~1(a,b) in the main text]. 
In Fig.~\ref{fig:Current-W}, we show the current versus time, for different interaction strengths.
We take $W=1$ (the blue dashed curve) as a reference. 
For vanishing or weak 1PU scattering, the packets pass through each other 
ballistically and the current switches sign every time the packets return to the $y$-edges. 
With stronger and stronger 1PU interaction, the bounce dynamics become more and more efficient and the current 
now switches sign each time the packets with opposite density collide on $y$-edges. 
This results in the frequency doubling of the current switching, also shown in Fig.~4(d) of the main text. 

The crossover from ballistic transport to retroreflection for oppositely charged packets
occurs for $W n_0 \sim 0.1$ (when $n_0/T_0 \ll 1$). This is 
evidenced by comparing Figs.~\ref{fig:Current-W} and \ref{fig:Current-W-n0=0.1}.